\documentclass[12pt]{article}
\usepackage{epsfig}
\usepackage{amssymb}

\newcommand{\be}{\begin{equation}}
\newcommand{\ee}{\end{equation}}
\newcommand{\bea}{\begin{eqnarray}}
\newcommand{\eea}{\end{eqnarray}}
\newcommand{\ov}{\overline}

\newcommand{\ba}{\begin{array}}
\newcommand{\ea}{\end{array}}

\begin{document}

\thispagestyle{empty}

\begin{flushright}
ITEP-PH-2/2001\\
hep-ph/0104048
\end{flushright}

\vspace{30mm}
\centerline{\LARGE\bf The check of QCD based on the {\boldmath $\tau$}-decay }
\centerline{\LARGE\bf  data analysis in the complex {\boldmath $q^2$}-plane }
\vspace{15mm}
\centerline{\large B.V. Geshkenbein, B.L. Ioffe and K.N. Zyablyuk}
\vspace{5mm}
\centerline{\tt geshken@vitep5.itep.ru, ioffe@vitep5.itep.ru, zyablyuk@heron.itep.ru}
\vspace{5mm}
\centerline{\it Institute of Theoretical and Experimental Physics,}
\centerline{\it B.Cheremushkinskaya 25, Moscow 117259, Russia}
\vspace{10mm}

\begin{abstract}
The thorough analysis of the ALEPH data \cite{ALEPH2} on hadronic $\tau$-decay is performed
in the framework of QCD. The perturbative calculations are performed in 3 and 4-loop approximations.
The terms of the operator product expansion (OPE) are accounted up to dimension $D=8$.
The value of the QCD coupling constant $\alpha_s(m_\tau^2)=0.355\pm 0.025$ was found from
hadronic branching ratio $R_\tau$. The $V+A$ and $V$ spectral function are analyzed
using analytical properties of polarization operators in the whole complex $q^2$-plane. Borel
sum rules in the complex $q^2$ plane along the rays, starting from the origin, are used.
It was demonstrated that QCD with OPE terms is in agreement with the data
for the coupling constant close to the lower error edge $\alpha_s(m_\tau^2)=0.330$.
The restriction on the value of the gluonic condensate was found
$\bigl<{\alpha_s\over\pi}G^2\bigr>=0.006\pm 0.012\,{\rm GeV}^2$. The analytical perturbative
QCD was compared with the data. It is demonstrated to be in strong contradiction with
experiment. The restrictions on the renormalon contribution were found. The instanton contributions
to the polarization operator are analyzed in various sum rules.
In Borel transformation they
appear to be small, but not in spectral moments sum rules.
\end{abstract}

\vspace{10mm}
\centerline{PACS: 13.35.D, 11.55.H, 12.38}

\newpage

\section{Introduction}

The high precision data on hadronic $\tau$-decay, obtained by ALEPH \cite{ALEPH2}, OPAL \cite{OPAL}
and CLEO \cite{CLEO} collaborations, namely the measurements of the total hadronic branching ratio
$R_\tau=B(\tau\to \nu_\tau+{\rm hadrons})/B(\tau\to e\bar{\nu}_e\nu_\tau)$, vector $V$ and axial $A$
spectral functions allow one to perform various tests of QCD at low energies: to determine
$\alpha_s(Q^2)$ at low $Q^2$, to check the operator product expansion (OPE) and to perform search
for other possible nonperturbative modifications of QCD --- renormalons, analytical $\alpha_s(Q^2)$,
instantons etc. An early attempt to check OPE in QCD based on $e^+e^-$ annihilation data has been
made by Eidelman, Vainstein and Kurdadze \cite{EVK} but the accuracy of the data at that
time was not good enough. Also the authors of \cite{EVK} took as granted that the QCD coupling
constant is rather small, $\Lambda^{(3)}_{QCD}$ (for 3 flavors) is about 100 MeV and neglected
higher order terms of perturbative series. Now it is common belief that $\alpha_s$ is much
larger and $\Lambda^{(3)} \sim 300-400 \, {\rm MeV}$ in 2--3 loop approximation. Therefore the problem
deserves reconsideration.

In the previous paper by two of us (B.I. and K.Z.) \cite{IZ} the difference of vector and axial current
correlators was analyzed using ALEPH data on $\tau$-decay \cite{ALEPH2}. The analytical properties
of the polarization operator in the whole complex $q^2$-plane were exploited and the vacuum
expectation values of dimension 6 and 8 operators (vacuum condensates) were found. Here we
consider $V+A$ correlator, where perturbative corrections are dominant.

Define the polarization operators of hadronic currents:
\be
\label{pol0}
\Pi_{\mu\nu}^J(q)\,=\,
i\!\int e^{iqx}\left<TJ_\mu(x) J_{\nu}(0)^\dagger\right> dx\,=\,\left(q_\mu q_\nu -g_{\mu\nu} q^2\right)
\Pi_J^{(1)}(q^2)+ q_\mu q_\nu \Pi_J^{(0)}(q^2)  \; ,
\ee
$$
{\rm where} \qquad J=V,A \, ;   \qquad
V_\mu=\bar{u}\gamma_\mu d  \; , \qquad
A_\mu=\bar{u}\gamma_\mu\gamma_5 d \; .
$$
The imaginary parts of the correlators are the so-called spectral functions ($s=q^2$),
\be
v_1/a_1(s)=2\pi\,{\rm Im}\,\Pi^{(1)}_{V/A}(s+i0) \; , \qquad
a_0(s)=2\pi\,{\rm Im}\,\Pi^{(0)}_{A}(s+i0) \; .
\ee
which have been measured from hadronic $\tau$-decays
for $0<s<m_\tau^2$.

The spin-1 parts $\Pi^{(1)}_V(q^2)$ and $\Pi^{(1)}_A(q^2)$ are
analytical functions in the complex $q^2$-plane with a cut along the right semiaxes starting from
the threshold of the lowest hadronic state: $4m_\pi^2$ for $\Pi^{(1)}_V$ and
$9m_\pi^2$ for $\Pi^{(1)}_A$. The latter has a kinematical pole at $q^2=0$. This is a
specific feature of QCD, which follows from the chiral symmetry in the limit of massless
$u,d$-quarks and its spontaneous violation. It can be easily shown \cite{GL} (see also \cite{IZ}),
that the kinematical pole arises from the pion contribution
to $\Pi^A_{\mu\nu}$, which is given by
\be
\label{pola3}
\Pi_{\mu\nu}^A(q)_\pi\,=\,-\,{f_\pi^2\over q^2}\left( \, q_\mu q_\nu - g_{\mu\nu}q^2\, \right)
-\,{m_\pi^2\over q^2}\, q_\mu q_\nu \, {f_\pi^2\over q^2- m_\pi^2}
\ee
where $f_\pi$ is the pion decay constant, $f_\pi=130.7 \; {\rm MeV}$ \cite{PDG}.

\section{Hadronic branching ratio and the value of $\alpha_s(m_\tau^2)$}

The total hadronic branching ratio into final state with zero strangeness is given by well known
expression, which can be written in the following form (see e.g.~\cite{P1}):
$$
R_{\tau,\, V+A} \, = \, {B(\tau\to \nu_\tau + {\rm hadrons}_{S=0})\over B(\tau\to \nu_\tau e\bar{\nu}_e)}
\hspace{75mm}
$$
\be
=\,6 |V_{ud}|^2 S_{EW} \int_0^{m_\tau^2}\!{ds\over m_\tau^2}
\left( 1-{s\over m_\tau^2} \right)^2\left[ \left( 1+2 {s\over m_\tau^2}\right)
( v_1+a_1+a_0)(s) - 2 {s\over m_\tau^2} a_0(s)\right]
\label{rtau2}
\ee
where $|V_{ud}|=0.9735\pm 0.0008$ \cite{PDG} is Cabbibo-Kabayashi-Maskava matrix element,
$S_{EW}=1.0194\pm 0.0040$ includes electroweak corrections \cite{MS}.
The spin-0 axial spectral function $a_0(s)$ is basically saturated by
$\tau\to \pi\nu_\tau$ channel and can be read off from (\ref{pola3}):
$a_0(s)=2\pi^2 f_\pi^2\delta (s-m_\pi^2)$. So the last term
in (\ref{rtau2}) gives small correction
\be
\Delta R_{\tau}^{(0)}=\,-\,24\,\pi^2\, {f_\pi^2 m_\pi^2\over m_\tau^4} \,=\,-0.008
\label{delr0}
\ee
The rest of (\ref{rtau2}) contains only the imaginary part of
$\Pi_{V+A}^{(1)}(s)+\Pi_A^{(0)}(s)$, for which the short notation $\Pi(s)$ will be used later on.
As follows from (\ref{pola3}), $\Pi_A^{(0)}(q^2)$ compensates the kinematical pole at $q^2=0$ in
$\Pi_A^{(1)}(q^2)$. So the combination $\Pi(q^2)$ has no kinematical poles and
is an analytical function of $q^2$ in the complex $q^2$-plane with a cut along the positive real axis.

The convenient way to calculate the $R_\tau$ in QCD or, turning the problem around, to find
$\alpha_s(m_\tau^2)$ from experimentally known $R_\tau$ is to transform the integral in  (\ref{rtau2})
to the integral over the contour in the complex $s$-plane going couterclockwise around the circle
$|s|=m_\tau^2$ \cite{B1}---\cite{DP1}:
\be
R_{\tau,\, V+A} = 6\pi i |V_{ud}|^2 S_{EW} \oint_{|s|=m_\tau^2}\!{ds\over m_\tau^2}
\left( 1-{s\over m_\tau^2} \right)^2 \left( 1+2 {s\over m_\tau^2}\right) \Pi (s) + \Delta R_\tau^{(0)}
\label{rtau3}
\ee
The polarization operator is given by the sum of perturbative and nonperturbative terms.
If we restrict ourselves by OPE terms, then
\be
\Pi(s) \, =\, -\,{1\over 2\pi^2} \, \ln{-s\over \mu^2} \,+\, \mbox{higher loops}\, +\, \sum_{n\ge 2}
{\left<O_{2n}\right>\over (-s)^n} \left( 1+ c_n {\alpha_s\over \pi} \right)
\label{polop1}
\ee

Consider at first the perturbative part. For its calculation it is convenient to use Adler function
which is perturbatively constructed as an expansion in coupling constant
\be
D(Q^2) \, \equiv \, - 2\pi^2 \,{ d\Pi(Q^2)\over d\ln{Q^2}} \,=\,\sum_{n\ge 0} K_n a^n
 \; , \qquad a\equiv {\alpha_s \over \pi}\; , \qquad  Q^2\equiv -s
\label{adler1}
\ee
which is known up to the 4-loop term in $\ov{\rm MS}$ renormalization scheme :
$K_0=K_1=1$ and $K_2=1.64$ \cite{CKT},
$K_3=6.37$ \cite{SS} for 3 flavors. The renormalization group equation
for $a(Q^2)$ reads:
\be
{d a \over d \ln{Q^2}} \, =\, -\beta(a) \,=\, - \sum_{n\ge 0} \beta_n a^{n+2}  \qquad  \Rightarrow
 \qquad \ln{Q^2\over \mu^2}\, = \,-\, \int_{a(\mu^2)}^{a(Q^2)} {da\over \beta(a)}
\label{rge}
\ee
In $\ov{\rm MS}$ scheme for 3 flavors $\beta_0=9/4$, $\beta_1=4$, $\beta_2=10.06$,
$\beta_3=47.23$ \cite{TVZ, RVL}. This allows us to get the perturbative contribution to the polarization
operator explicitly at any order of perturbation theory:
\be
\Pi(Q^2)\,-\,\Pi(\mu^2)\,=\,{1\over 2\pi^2} \int_{a(\mu^2)}^{a(Q^2)} D(a) {da\over \beta(a)}
\label{pi3}
\ee
Let us put $\mu^2= m_\tau^2$ and choose some value $a(m_\tau^2)$. From (\ref{rge}) we can
find $a(Q^2)$ for any $Q^2$ and by analytical  continuation at any $s$. Computing the integral
(\ref{pi3}) it is possible to find the perturbative part of $\Pi(s)$ as a function of $a(s)$ in the whole
complex $s$-plane. The substitution of $\Pi(s)$ into (\ref{rtau3}) gives (up to the power corrections)
the dependence of $R_\tau$ on $a(m_\tau^2)$. It must be stressed, that in this calculation
no expansion in inverse powers of $\ln{Q^2}$ is performed: only the validity of expansion
series in (\ref{adler1}) and (\ref{rge}) is assumed\footnote{Such way of calculation in \cite{ALEPH2}
was called contour-improved fixed-order perturbation theory.}.
Such representation has a serious advantage:
on the right semiaxes, i.e. in the physical region, there is no expansion in $\pi/\ln{(Q^2/\Lambda^2)}$,
which is not small at intermediate $Q^2$. For instance in the next to leading order
\be
2\pi \,{\rm Im} \, \Pi (s+i0) \,=\,1+{1\over \pi\beta_0}\left[ {\pi\over 2}- \arctan\left( {1\over\pi}
\ln{s\over\Lambda^2} \right) \right]
\label{impi1}
\ee
$$
\mbox{instead of }  \qquad
2\pi \,{\rm Im} \, \Pi (s+i0) \,=\,1+{1\over \beta_0\ln{(s/\Lambda^2)}} \hspace{3cm}
$$
which would follow in case of small $\pi/\ln{(s/\Lambda^2)}$. (Eq.~(\ref{impi1}) was first obtained
in \cite{ShrShr}, the systematical method of analytical continuation from space-like to time-like
region with summation of $\pi^2$-terms was suggested in \cite{Rad} and developed in \cite{Piv}.)
In the higher order, where $a(s)$ cannot be
expressed via $\ln{(s/\Lambda^2)}$ in terms of elementary functions,
this analysis is performed numerically.

\begin{figure}[tb]
\hspace{20mm}
\epsfig{file=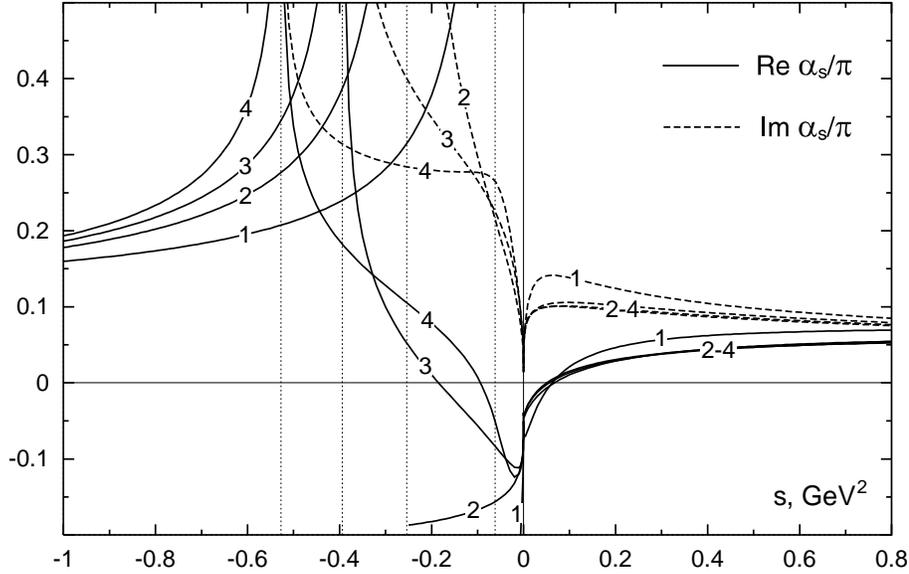, width=120mm}
\caption{Real and imaginary parts of $\alpha_{\ov{\rm MS}}(s)/\pi$  as
exact numerical solution of RG equation (\ref{rge}) on real axes for different number of loops.
The initial condition is chosen
$\alpha_s=0.355$ at $s=-m_\tau^2$, $N_f=3$. Vertical dotted lines display the
position of the unphysical singularity at $s=-Q_0^2$ for each approximation ($4\to 1$ from left to right).}
\label{alpha}
\end{figure}

It is well known, that in the 1-loop approximation of $\beta$ function the coupling
$a(Q^2)$ has an infrared pole at some $Q^2=Q^2_0$ (in some conventions coinciding with
$\Lambda^2$).
In the $n$-loop approximation ($n>1$) instead of pole a branch cut appears
with a singularity $\sim (1-Q^2/Q^2_0)^{-1/n}$. The position of the singularity is given by
\be
\ln{Q_0^2\over \mu^2}\, = \,-\, \int_{a(\mu^2)}^\infty {da\over \beta(a)}
\ee
Near the singularity the last term in the expansion of $\beta(a)$ (\ref{rge}) dominates and gives the
aforementioned behavior. To illustrate the behavior of the running coupling constant,
we plotted real and imaginary part of $\alpha_s/\pi$ for $n=1,2,3,4$-loop $\beta$-function in Fig
\ref{alpha}. It demonstrates, that for real positive $s$ the difference between various approximations
is almost unnoticable beyond 2-nd loop and the expansion in inverse $|\ln{(s/\Lambda^2)}-i\pi |$
works well. At the same time the behavior in  the unphysical cut strongly depends on the number of
loops and cannot be described by some simple approximation. Only at $s<1\,{\rm GeV}^2$ 2--4 loop
calculations more or less coincide.

Let us turn now to OPE terms in (\ref{polop1}).  The contribution of the operators
up to dimension 8 have been computed theoretically:
\bea
\sum_{n\ge 2} {\left<O_{2n}\right>\over (-s)^n} \left( 1+ c_n {\alpha_s\over \pi} \right) & = &
 {\alpha_s\over 6 \pi\, Q^4} \left< G_{\mu\nu}^aG_{\mu\nu}^a\right>\left( 1+
{7\over 6} {\alpha_s\over \pi} \right) \nonumber \\
 &+ & {128\over 81\, Q^6}\, \pi\alpha_s \left<\bar{q}q\right>^2 \left[ 1 + \left({29\over 24} +
{17\over 18}\ln{Q^2\over \mu^2}  \right){\alpha_s\over \pi}
 \right] + {\left<O_8\right>\over Q^8}  \label{ope}
\eea
The contribution of $D=2$ operator due to nonzero quark masses $m_{u, d}$ is negligible and
omitted here. We have also neglected the $D=4$ quark condensate
$2(m_u+m_d)\left<\bar{q}q\right>$
which is an order of magnitude less than the gluonic condensate. The coefficients in front of
$D=4,6$ operators have been computed in \cite{SVZ}, hereafter cited as SVZ.
The $\alpha_s$-correction to the $D=4$ operator
were found in \cite{CGS}; $\alpha_s$-corrections to $D=6$ operator were calculated
\cite{AC}; ambiguities among them were also discussed there.

Few comments about the operator $O_6$ are in order.
In nonfactorized form without $\alpha_s$-corrections it looks as follows \cite{SVZ}:
\bea
\left<O_6\right> & = & -2\pi\alpha_s \Biggl<
(\bar{u}\gamma_\mu\lambda^a d)(\bar{d}\gamma_\mu \lambda^a u) +
(\bar{u}\gamma_5\gamma_\mu\lambda^a d)(\bar{d}\gamma_5\gamma_\mu \lambda^a u) \nonumber \\
 & & \hspace{13mm} +\,{2\over 9} (\bar{u}\gamma_\mu\lambda^a u +\bar{d}\gamma_\mu\lambda^a d)
\sum_{u,d,s} (\bar{q}\gamma_\mu \lambda^a q) \Biggr>   \label{o6nf}
\eea
After factorization three terms in (\ref{o6nf}) give the following contributions:
\be
\left<O_6\right> \,=\,4\pi\alpha_s\left<\bar{q}q\right>^2 \left( 1-{1\over N_c^2} \right)
\left( 1-1+{4\over 9} \right) \label{o6f}
\ee
where $N_c$ is the number of colors.  SVZ assumed that the accuracy of the factorization
procedure is of order $N_c^{-2} \sim 10\%$ in case of $V$-correlator, where the coefficient in
the second brackets in (\ref{o6f}) is equal to $-7/9$.
Remind that in $V-A$ correlator the first term has opposite sign
and the third term is absent, so the accuracy of the factorized operator $O_6^{V-A}$
is at least not worse, than in $V$ case. On the other hand in the $V+A$ correlator
two comparatively large terms cancel each other under the factorization assumption in (\ref{o6f}).
Consequently the accuracy of the formula (\ref{o6f}) for the operator $O_6$ is less, perhaps
$20-30\%$. Large $\alpha_s$-corrections to all independent $D=6$ operators \cite{AC}
can only increase the errors.

The numerical value of $D=6$ operator can be estimated, for instance, with help of our previous
analysis of $V-A$ sum rules \cite{IZ}:
\be
\bigl<O_6^{V-A}\bigr>\,=\,-\,{64\over 9} \,\pi\alpha_s \left<\bar{q}q\right>^2 \times 1.3
\,=\,-\,(6.8\pm 2.1)\times 10^{-3} \,{\rm GeV}^6
\label{o6vma}
\ee
The coefficient $1.3$ stands for the $\alpha_s$-corrections.  We find:
\be
\left<O_6\right>\,=\,(1.3\pm 0.5)\times 10^{-3} \, {\rm GeV}^6 \; ,
\label{o6num}
\ee

The dimension 8 operators come from many different diagrams, which
can be labeled by the number of quarks in vacuum. The purely gluonic
condensates are suppressed by the loop factor $\sim \alpha_s/\pi$ and are neglected on this ground.
The 4-quark operators, computed in \cite{DS, GP} and \cite{IZ},
 vanish in the sum $V+A$ after factorization. The uncertainty of this cancellation
can be estimated as $\sim 10\%$ of $O_8^{V-A}$, which is about $10^{-3}\,{\rm GeV}^8$.
The 2-quark operators have the same sign in $V$ and $A$ correlators. They have been computed
in \cite{BG} (we have performed the calculation independently to confirm this result)
and can be written in the following form:
\bea
\bigl<O_8^{(2)}\bigr> & = & {2\over 9}\,
\Biggl< \,2i\,\bar{u}\gamma^\alpha  \{ G^2_{(\alpha \beta )}, D^\beta  \} u
\,-\,\bar{u}\gamma^\alpha \gamma^5 \{ (\tilde{G}G)_{\alpha \beta }, D^\beta \} u
\,+\,{i\over 4}\,\bar{u}\gamma^\gamma\! \left[ G^{\alpha \beta },(D_\gamma G_{\alpha \beta})  \right]\! u
\nonumber \\
 & &
\left.
\,+\,{1\over 2}\,\bar{u}(D^2\hat{J}) u
\,-\,i\bar{u}\gamma^\alpha\! \left[ G_{\alpha \beta }, J^\beta  \right]\! u \, \right> + \, (u\to d)
\label{o8}
\eea
where $J_\mu = D_\nu G_{\mu\nu}=\pi\alpha_s \lambda^a \sum_q (\bar{q}\gamma_\mu\lambda^a q)$.
The last two terms can be factorized and brought to the form
$\pi\alpha_s\!\left<\bar{q}q\right>\!\bigl<\bar{q}\hat{G}q\bigr>$.
However the leading in the number of colors $N_c^0$ terms
cancel each other and only the terms $\sim N_c^{-2}$ are left. It has been shown in \cite{IZ},
that the factorization of $D=8$ operators is not unambiguous at this level of accuracy.
Taking the value of the operator $\bigl<\bar{q}\hat{G}q\bigr>$ from \cite{BI, IZ}, we may
estimate the upper limit of the operator (\ref{o8}) as
$|\bigl<O_8^{(2)}\bigr>|<10^{-4} \,{\rm GeV}^8$, which is tiny.
So, for the upper limit of the total $D=8$ operator we shall use
the estimation $|\bigl<O_8\bigr>|<10^{-3} \,{\rm GeV}^8$.

It is worth mentioned, that the $D=6,8$ operators in $V+A$ polarization function are much smaller,
than in $V$ or $A$ separately.

We are now in position to calculate $\alpha_s(m_\tau^2)$ from the experiment. We take the most recent
data on the total hadronic decay ratio $R_\tau$ \cite{PDG} and the ratio of decays with odd
number of strange mesons $\tau^-\to X(S=-1)\nu_\tau$ \cite{ALEPH3, OPAL3}:
\be
R_\tau\,=\,3.636 \pm 0.021   \; , \qquad  R_{\tau, S}\,=\,0.161\pm 0.007
\ee
In our analysis we subtract $R_{\tau, S}$ to avoid the interference with additional parameters,
in particular the mass of $s$-quark. One obtains
\be
\label{rtvpa}
R_{\tau, V+A}  = 3|V_{ud}|^2 S_{EW}\left( \,1\,+\,\delta_{EW}'\,+\,\delta^{(0)}\,+\,\delta^{(6)}_{V+A}
\, \right) +\Delta R^{(0)}\,=\, 3.475 \pm 0.022
\ee
where $\Delta R^{(0)}_\tau$ is given by (\ref{delr0}). We use conventional in $\tau$-literature
notations of fractional corrections $\delta$. The electromagnetic correction is
$\delta_{EW}'={5\over 12\pi}\alpha_{em}(m_\tau^2)=0.001$ \cite{BL}, the $D=6$ operator
correction $\delta^{(6)}_{V+A}=-(5\pm 2) \times 10^{-3}$ as follows from our analysis, in agreement
with the estimation obtained in \cite{BNP}. From (\ref{rtvpa}) we separate out the perturbative
correction:
\be
1\,+\,\delta^{(0)}\,=\,1.206\pm 0.010
\label{delta0}
\ee
All errors in here are added in quadratures (perhaps, such procedure underestimates the total
error, may be by a factor 2).

\begin{figure}[tb]
\hspace{25mm}
\epsfig{file=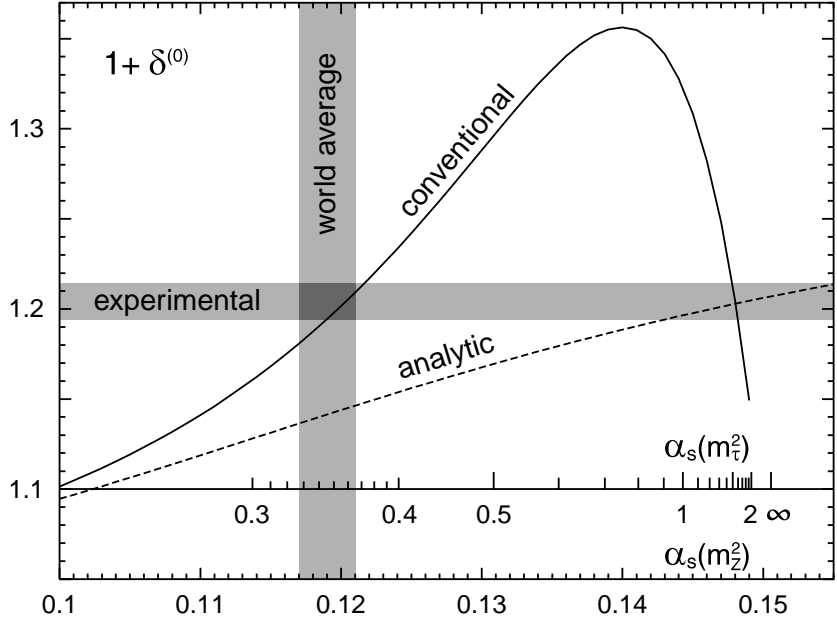, width=110mm}
\caption{Perturbative fractional correction $\delta^{(0)}$ versus
$\alpha_s(m_\tau^2)$ and $\alpha_s(m_Z^2)$ in conventional and analytical approach
in 3-loop approximation. In the width of the experimental strip the theoretical uncertainty of
the operator $\left<O_6\right>$ is included.}
\label{delta0fig}
\end{figure}

The calculation of $\alpha_s(m_\tau^2)$ corresponding to $\delta^{(0)}$ were performed according to
the method described above. 
The dependence of $1+\delta^{(0)}$ on $\alpha_s(m_\tau^2)$ (and on
$\alpha(M_Z^2)$, to compare with other data) for 3-loop $\beta$-function and 3-loop Adler
function is shown of Fig \ref{delta0fig}. It follows from Fig \ref{delta0fig}:
\be
\alpha_s(m_\tau^2)\,=\,0.355\pm 0.025
\label{atau}
\ee
The estimation of the error in (\ref{atau}) was done with care. 
Because of asymptotic character of perturbative series (\ref{adler1}) and (\ref{rge})
the higher loop contribution could be as large as the contribution of the last terms, namely 
$K_3 a^3$ and $\beta_2 a^4$. They result to uncertainty $0.015-0.20$ in $\alpha_s(m_\tau^2)$,
depending on its central value. Taking into account the uncertainty (\ref{delta0}) in $\delta^{(0)}$,
we obtain the error in (\ref{atau}). Furthermore, we have performed 2 and 4-loop calculation of
$\alpha_s(m_\tau^2)$. The unknown 4-loop coefficient in Adler function 
(\ref{adler1}) was taken equal to $K_4=50$ (cf. its estimations \cite{KS}). For each given 
$\delta^{(0)}$ the 4-loop $\alpha_s(m_\tau^2)$ is by $0.005$ lower than 3-loop value,
while 2-loop $\alpha_s(m_\tau^2)$ is by $0.02$ higher. 
These results are within the error range (\ref{atau}).
If some nonperturbative terms beyond OPE exist (e.g. instantons), they would also contribute
to the error in (\ref{atau}).
In section 4 it will be shown, that the value $\alpha_s(m_\tau^2)$ close to the lower limit of
(\ref{atau}) satisfies sum rules at low $Q^2$ much better.

\section{$\alpha_s(m_\tau^2)$ and analytical QCD}

Shirkov and Solovtsov \cite{ShSol1}
forwarded the idea of analytical QCD. According to it the coupling constant $\alpha_s(Q^2)$
is calculated by renormalization group in the space-like region $Q^2>0$. Then, by analytical
continuation to $s=-Q^2>0$, $\alpha_s(s)$ was found, in particular its imaginary part
${\rm Im}\,\alpha_s(s)$ on the right semiaxes. It was assumed, that $\alpha_s(s)$
is an analytical function in the complex $s$-plane with a cut along the right semiaxes $0\le s<\infty$.
The analytical $\alpha_s(s)_{\rm an}$
is then defined in the whole complex $s$-plane by dispersion relation:
\be
\alpha_s(s)_{\rm an} \,=\, {1\over \pi} \, \int_0^\infty {{ \rm Im}\, \alpha_s(s') \over s-s'} \, ds'
\ee
Since the lower limit in this integral is put to zero, $\alpha_s(s)_{\rm an}$ indeed has no
unphysical singularities (poles, cuts etc) at $Q^2>0$. The idea of analytical QCD has been
developed in many papers, see e.g. \cite{ShSol2} and for review \cite{ShSol3}.
In particular the calculations of
$\alpha_s(m_\tau^2)_{\rm an}$ from $\tau$-decay data were performed within the framework of
analytical QCD in \cite{MSS}.

A related approach was suggested by two of us (B.G. and B.I.) in \cite{GI}. We started from
well-known theorem, that the polarization operator for $e^+e^-$ annihilation $\Pi(s)$ is
an analytical function of $s$ in complex $s$-plane with a cut along positive semiaxes and
\underline{assumed} that these analytical properties take place separately for perturbative and
nonperturbative parts of $\Pi(s)$. In the first order of $\alpha_s$ this hypothesis is equivalent
to analytical QCD while in higher orders it may be more general.

Let us calculate $\alpha_s(m_\tau^2)_{\rm an}$ in the framework of analytical QCD from the
same experimental data, i.e. $\delta^{(0)}$ given by (\ref{delta0}). The only (but important)
difference from the previous calculation is the following. The coupling $\alpha_s(s)_{\rm an}$ is
an analytical function of $s$ with a cut from $s=0$ to $s=\infty$. Consequently the contour
integral in (\ref{rtau3}) is now equal to the original integral (\ref{rtau2}) with ${\rm Im}\, \Pi(s)$
over real positive axes. In the previous calculation, if such transformation is performed, the integral
would run from $s=-Q_0^2$ to $m_\tau^2$. Qualitatively it leads to much smaller $R_\tau$ in
analytical QCD than in conventional approach with the same $\alpha_s(m_\tau^2)$, or vice versa,
the same $R_\tau$ corresponds to much larger $\alpha_s(s)_{\rm an}$.
Direct numerical calculation confirms this expectation. The dependence of $1+\delta^{(0)}$
versus $\alpha_s(s)_{\rm an}$ is also displayed in Fig \ref{delta0fig}.
It is seen, that in order to get experimental
value of $\delta^{(0)}$ in analytical QCD one should take $\alpha_s(m_\tau^2)_{\rm an}\approx 1.5 - 2.0$,
which corresponds to $\alpha_s(m_Z^2)\approx 0.15$ in strong contradiction with
the world average $\alpha_s(m_Z^2) = 0.119\pm 0.002$ \cite{PDG}. (The previous calculation of
$\tau$-decay \cite{MSS}, performed with less certainty, demonstrated the same trend: in particular
$\Lambda^{(3)}=700 - 900 \, {\rm MeV}$, much larger, than in standard calculations.)

It the recent paper \cite{MSS2} an attempt was made to save the analytical QCD in case of vector
polarization operator and to obtain the agreement with ALEPH data on vector Adler $D$-function
by assuming large quark masses $m_u=m_d=250\,{\rm MeV}$ and some form of Coulomb-like
quark-antiquark interaction. This hypothesis, however, is in strong contradiction with all results
following from well established partial conservation of axial current (PCAC) and chiral theory.
For example, Gell-Mann-Oakes-Renner relation and $K/\pi$ mass ratios would be violated by
an order of magnitude, Goldberg-Treitman relation cannot be proved etc. Also many
sum rules for $V-A$ polarization operator would disagree with the data.

Therefore we come to the conclusion, that analytical QCD in any form,
\cite{ShSol1} or \cite{GI} is in strong contradiction
with experiment and must be abandoned.

\section{Check of QCD at low $Q^2$ for $V+A$ correlators by using the sum rules}

Let us turn now to study of the $V+A$ correlator in the domain of low $Q^2$, where the OPE terms
play much more essential role, than in the determination of $R_\tau$. A general remark is in order
here. As was mentioned in \cite{NSVZ} and stressed recently by Shifman \cite{S1}, the condensates
cannot be defined in rigorous way, because there is some arbitrariness in the separation of their
contributions from perturbative part. Usually \cite{NSVZ, S1} they are defined by introduction
of some normalization point $\mu^2$ with the magnitude of few $\Lambda^2$.  The integration
over momenta in the domain below $\mu^2$ is addressed to condensates, above $\mu^2$ --- to
perturbation theory. In such formulation the condensates are $\mu$-dependent
$\left<O_D\right>=\left<O_D\right>_\mu$ and, strictly speaking, they also depend
on the way how the infrared cut-off $\mu^2$ is introduced.
The problem becomes more severe when the perturbative expansion
is performed up to higher order terms and the calculation pretends on high precision.
Mention, that this remark does not refer to chirality violating condensates,
because perturbative terms do not contribute to chirality violating 
structures. For this reason, in principle, chirality violating condensates,
e.g.~$\bigl<0|\bar{q}q|0\bigr>$, can be determined with higher precission,
than chirality conserving ones.
 Here we use the definition of condensates, which can be 
called $n$-loop condensates.  As was formulated in Section 2, we treat the 
renormalization group equation (\ref{rge}) and the equation for polarization 
operator (\ref{pi3}) in $n$-loop approximation as exact ones; the expansion 
in inverse logarithms is not performed. Specific values of condensates are 
referred to such procedure. Of course, their numerical values depend on the 
accounted number of loops; that is why the condensates, defined in this way, 
are called $n$-loop condensates.

\begin{figure}[tb]
\hspace{35mm}
\epsfig{file=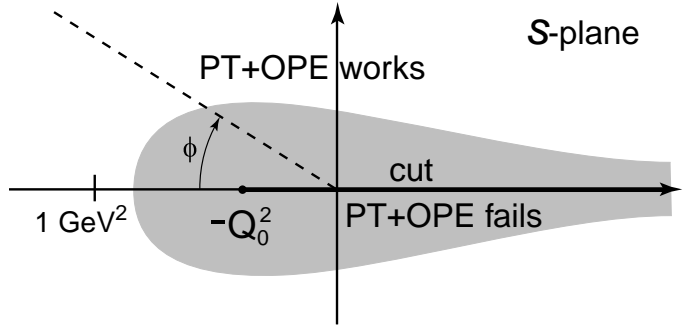, width=90mm}
\caption{Region of validity of the perturbation theory (PT) and operator product expansion (OPE)}
\label{pt_ope}
\end{figure}

Consider the polarization operator $\Pi=\Pi_{V+A}^{(1)}+\Pi_A^{(0)}$, defined in (\ref{pol0})
and its imaginary part
\be
\omega(s)\,=\,v_1(s)\,+\,a_1(s)\,+\,a_0(s)\,=\,2\pi\, {\rm Im}\,\Pi (s+i0)
\ee
In parton model $\omega(s) \to 1$ at $s\to \infty$.
Any sum rule can be written in the following form:
\be
\int_0^{s_0} f(s) \, \omega_{\rm exp}(s)\,ds \,=\,i\pi \oint f(s)\, \Pi_{\rm theor}(s) \,ds
\label{sr}
\ee
where $f(s)$ is some analytical in the integration region function.
In what follows we use $\omega_{\rm exp}(s)$, obtained from $\tau$-decay invariant
mass spectra published in \cite{ALEPH2} for $0<s<m_\tau^2$ with step
$ds=0.05 \, {\rm GeV}^2$. The experimental
error of the integral (\ref{sr}) is computed as the double integral with the covariance matrix
$\ov{\omega(s)\omega(s')}-\ov{\omega}(s)\ov{\omega}(s')$,
which also can be obtained from the data available in \cite{ALEPH2}.
In the theoretical integral in (\ref{sr})
the contour goes from $s_0+i0$ to $s_0-i0$ counterclockwise around
all poles and cuts of theoretical correlator $\Pi(s)$, see Fig \ref{pt_ope}.
Because of Cauchy theorem the
unphysical cut must be inside the integration contour.

The choice of the function $f(s)$ in (\ref{sr}) is actually a matter of taste. At first let us consider
usual Borel transformation:
\be
B_{\rm exp}(M^2) \,=\, \int_0^{m_\tau^2} e^{-s/M^2} \omega_{\rm exp} (s) \, {ds\over M^2} \, = \,
B_{\rm pt}(M^2) \,+\,2\pi^2 \sum_n {\bigl<O_{2n}\bigr>\over (n-1)! \, M^{2n}}
\label{bor0}
\ee
We separated out the purely perturbative contribution $B_{\rm pt}$, which is computed
numerically according to (\ref{sr}) and (\ref{adler1}--\ref{pi3}). Remind that Borel transformation
improves the convergence of OPE series because of the factors $1/(n-1)!$ in front of
operators and suppresses the contribution of high-energy tail, where the experimental error is
large. But it does not suppress the unphysical perturbative cut, the main source of the error in this
approach, even increase it since $e^{-s/M^2}>1$ for $s<0$.
So the perturbative part $B_{\rm pt}(M^2)$ can be reliably calculated
only for $M^2 \approx 0.8-1\, {\rm GeV}^2$ and higher; below this value the influence of the
unphysical cut is out of control.

\begin{figure}[tb]
\hspace{25mm}
\epsfig{file=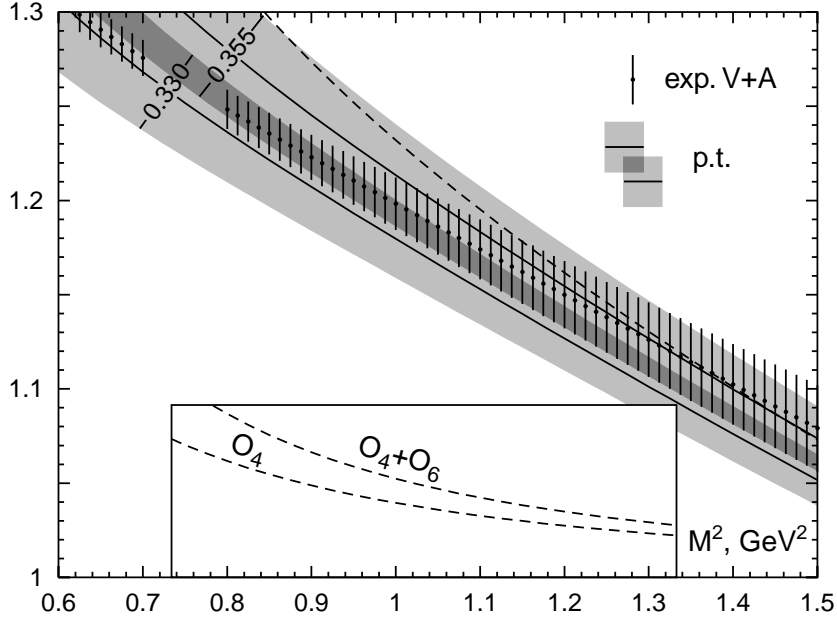, width=110mm}
\caption{Borel transformation (\ref{bor0}) $B_{\rm exp}(M^2)$ and $B_{\rm pt}(M^2)$ for
$\alpha_s(m_\tau^2)=0.355$ and $0.330$. The dashed line displays the OPE contribution added
to the $0.330$-perturbative curve. The contribution of the operators $D=4$
(standard SVZ value) and $D=6$
(central value of (\ref{o6num})) with respect to 1 are shown separately in the box.}
\label{bor0fig}
\end{figure}

Both $B_{\rm exp}$ and $B_{\rm pt}$
in 3-loop approximation for $\alpha_s(m_\tau^2)=0.355$ and $0.330$ are shown in Fig \ref{bor0fig}.
The shaded areas display the theoretical error. They are taken equal to the contribution of
the last term in the perturbative Adler function expansion $K_3 a^3$ (\ref{adler1}).
We have also performed the calculation with 4-loop $\beta$-function and $K_4=50\pm 50$, but
the result is very close to the 3-loop one, since positive contribution of the term $K_4 a^4$
compensates small decrease in the coupling $a$. Since this result is observed by us in many
other sum rules, we shall not give the 4-loop calculations later on and estimate the theoretical
error for any given $a(m_\tau^2)$ as the contribution of $K_3a^3$.

As follows from the analysis in Section 2, for $M^2>1\,{\rm GeV}^2$
the contribution of $D=6,8$ operators to the Borel transform  (\ref{bor0}) is small in $V+A$ channel,
 while the contribution of the $D=4$ condensate must be positive (we assume
$\alpha_s$-corrections included in the operators $\bigl<O_{2n}\bigr>$ in
(\ref{bor0}) and later). So the theoretical curve must go below experimental one. The result
shown in Fig \ref{bor0fig} is in favor of lower value of the coupling constant $\alpha_s(m_\tau^2)=0.33$.
Literally the theoretical curve (perturbative at $\alpha_s(m_\tau^2)=0.33$ plus the contribution of $O_4$
and $O_6$ operators) agrees with experiment starting from $M^2=1.1\,{\rm GeV}^2$. If the
uncertainties in perturbative contributions are taken into account (shaded area in Fig \ref{bor0fig})
the agreement may start earlier, at $M^2=1\,{\rm GeV}^2$.

The Borel transformation on Fig \ref{bor0fig} includes the contributions of different operators.
Although it is difficult to separate the perturbative part from the OPE one, the contributions
of different operators can be separated from each other.  One way is to differentiate
the Borel transformation by $M^2$. This however leads to the certain loss in the accuracy of
the experimental integral, since the growing power term $\sim s^n$ appears in the integral.
So we apply the method used in \cite{IZ} for $V-A$ sum rules, namely the Borel transformation
in complex $M^2$ plane.

\begin{figure}[tb]
\epsfig{file=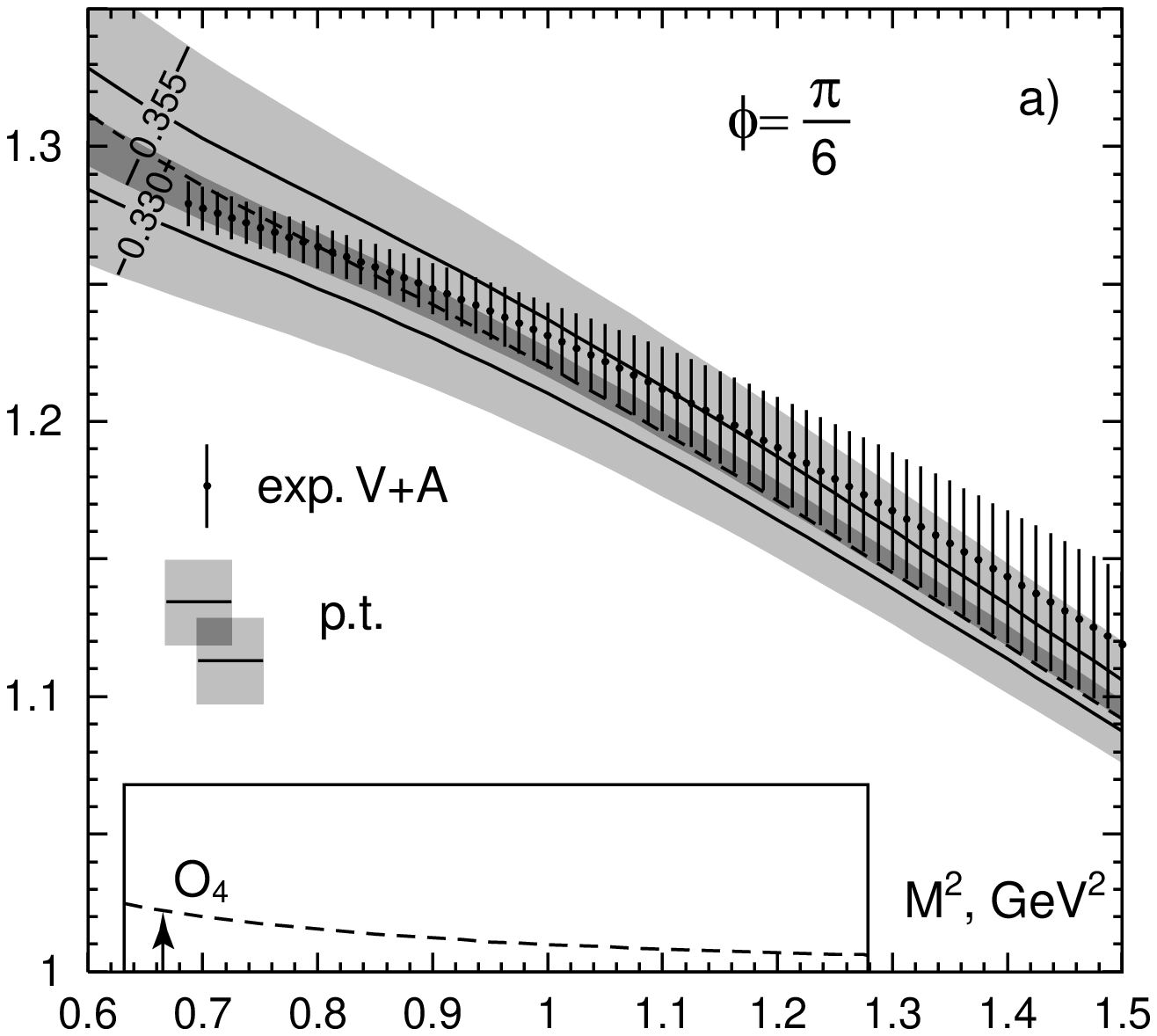, width=81mm}
\epsfig{file=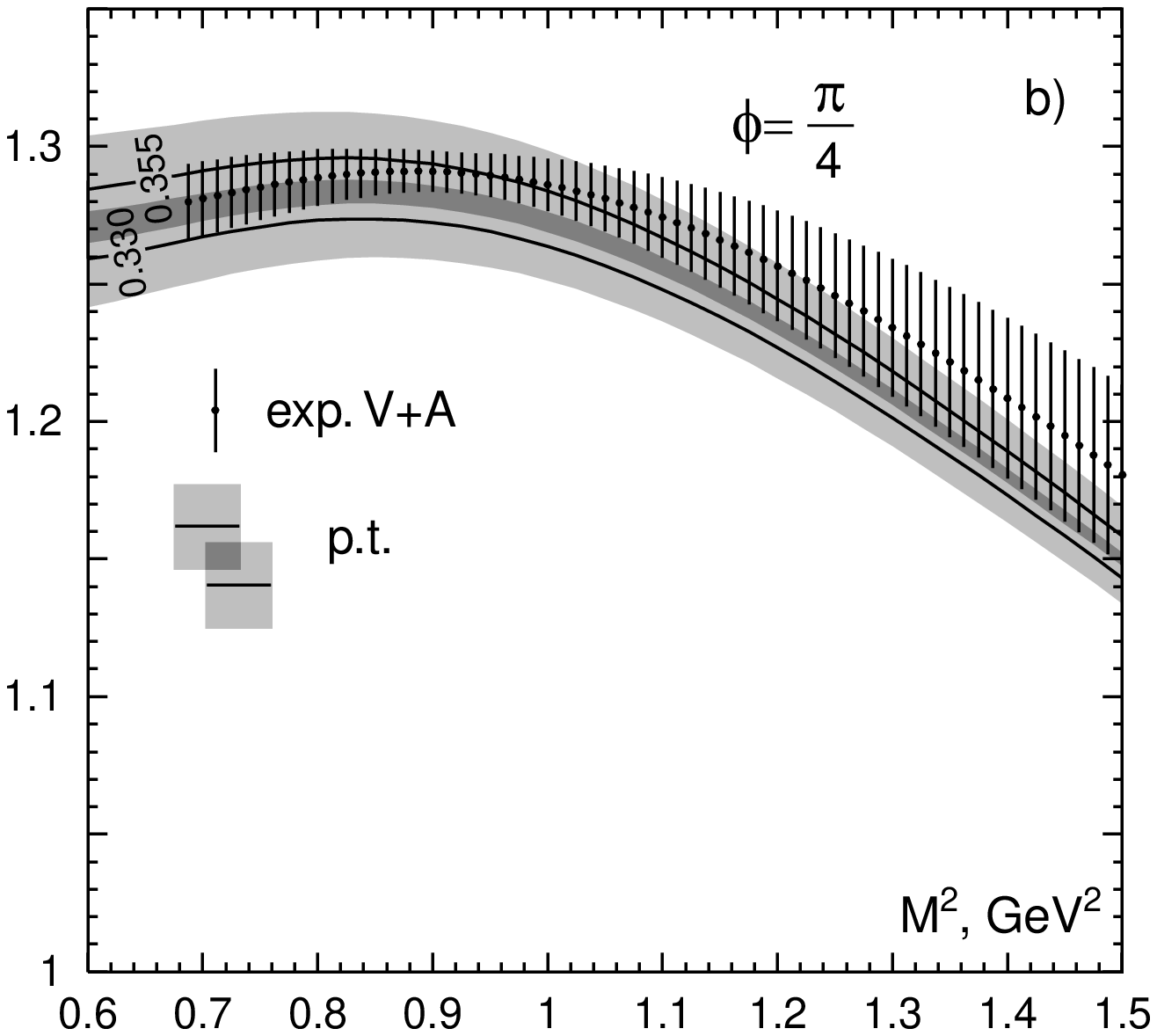, width=81mm}
\caption{Real part of the Borel transform (\ref{bor0}) along the rays at the angles $\phi=\pi/6$
and $\pi/4$ to the real axes. The dash line corresponds to the gluonic condensate given by
the central value of (\ref{condnum}).}
\label{set1819fig}
\end{figure}

Let us consider the Borel transform $B(M^2)$ (\ref{bor0}) at some complex $M^2=M_0^2e^{i\phi}$,
$0<\phi<\pi/2$. If the phase $\phi$ is taken close to $\pi/2$ then the contribution of the
high-energy tail becomes high. So we restrict ourselves by the values $\phi\le\pi/4$ for
the exponent to be decreasing enough.  The real part of the Borel transform at $\phi=\pi/6$
does not contain the $D=6$ operator:
\be
{\rm Re} \, B_{\rm exp} ( M^2e^{i\pi/6})\,=\,{\rm Re} \, B_{\rm pt} ( M^2e^{i\pi/6})
 \,+ \,\pi^2 \,{\bigl<O_4\bigr>\over M^4}
\label{set18}
\ee
The contribution of $\bigl<O_8\bigr>$ is less than $0.5\%$ to the perturbative term and neglected here.
The results are shown in Fig \ref{set1819fig}a.
Again it is still difficult to accommodate positive value of the gluonic condensate to the coupling
$\alpha_s(m_\tau^2)=0.355$ and higher. If we accept the lower value of $\alpha_s(m_\tau^2)$,
we get the following restriction on the
value of the gluonic condensate:
\be
\left< {\alpha_s\over\pi} \,G_{\mu\nu}^a G_{\mu\nu}^a \right> \, = \, 0.006\pm 0.012 \, {\rm GeV}^4
\; , \qquad \alpha_s(m_\tau^2)=0.330 \quad {\rm and} \quad M^2 > 0.8 \, {\rm GeV}^2
\label{condnum}
\ee
The theoretical and experimental errors are added together in (\ref{condnum}).

The real part of the Borel transform at $\phi=\pi/4$ does not contain the $D=4$ operator:
\be
{\rm Re} \, B_{\rm exp} ( M^2e^{i\pi/4})\,=\,{\rm Re} \, B_{\rm pt} ( M^2e^{i\pi/4})
 \,- \,\pi^2 \,{\bigl<O_6\bigr>\over \sqrt{2}M^6}
\label{set19}
\ee
The results are shown in Fig \ref{set1819fig}b.
The perturbative curve at $\alpha_s=0.330$ is below the data. If we would take this curve
as an exact one, without accounting the perturbative errors, then from (\ref{set19}) we would conclude,
that $\left<O_6\right> <0$, which in some contradiction with (\ref{ope}, \ref{o6num}). 
However the account for the perturbative errors makes the situation different, but uncertain. 
Since the value of the $\left<O_6\right>$  contribution to (\ref{set19}) is very small
\be
\pi^2 {\bigl<O_6\bigr> \over \sqrt{2} M^6}= (0.9 \pm 0.4)\times 10^{-2} \, {{\rm GeV}^6\over M^6}
\ee
then by accounting the perturbative errors it is possible to satisfy the sum rule (\ref{set19})
at positive $\left<O_6\right>$ starting from $M^2 >0.8\, {\rm GeV}^2$. (In the narrow region
near $M^2=0.9\,{\rm GeV}^2$ the theoretical curve goes out of the data on $1.5-2$ experimental
error, but we do not consider this as a serious contradiction.) Unfortunately no definite conclusion
about the value of $O_6$ can be done from the Fig \ref{set1819fig}b.
The only statement is that its value
cannot exceed (\ref{o6num}) and probably is on the lower border of error.

\section{Correlator of vector currents}

Previously we considered the $V+A$ correlators where the power corrections are small.
Instead one could take pure vector current, (vector spectral function was published
by ALEPH in \cite{ALEPH1}). This doesn't give us any new information with the $\tau$-decay data,
since $V-A$ correlators have already been analyzed in \cite{IZ}.
Moreover the accuracy of the vector current spectral  function is less, than $V+A$, since
 both currents are mixed in some channels with $K$-mesons and the number of events is
twice less.

However the analysis of the vector current correlator is important since it can also be performed
with the experimental data on $e^+e^-$ annihilation. The imaginary part of the
electromagnetic current correlator, measured here, is related to the charged current correlator
(\ref{pol0}) by the isotopic symmetry. The statistical error in $e^+e^-$ experiments is less
than in $\tau$-decays because of significantly larger number of events. So it would be
interesting to perform similar analysis with $e^+e^-$ data, which is a matter for separate
research.

\begin{figure}
\hspace{30mm}
\epsfig{file=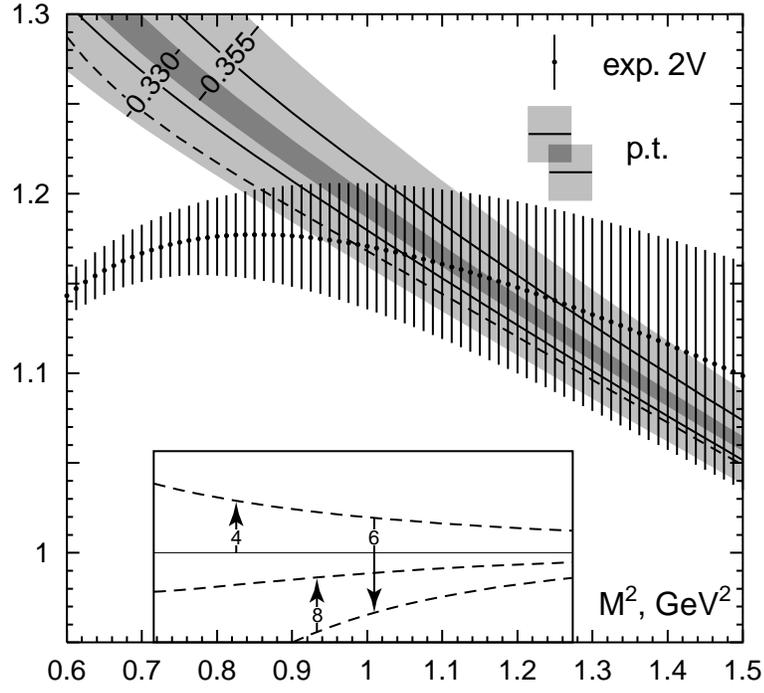, width=100mm}
\caption{Borel transformation for vector currents.}
\label{borvfig}
\end{figure}
\begin{figure}
\epsfig{file=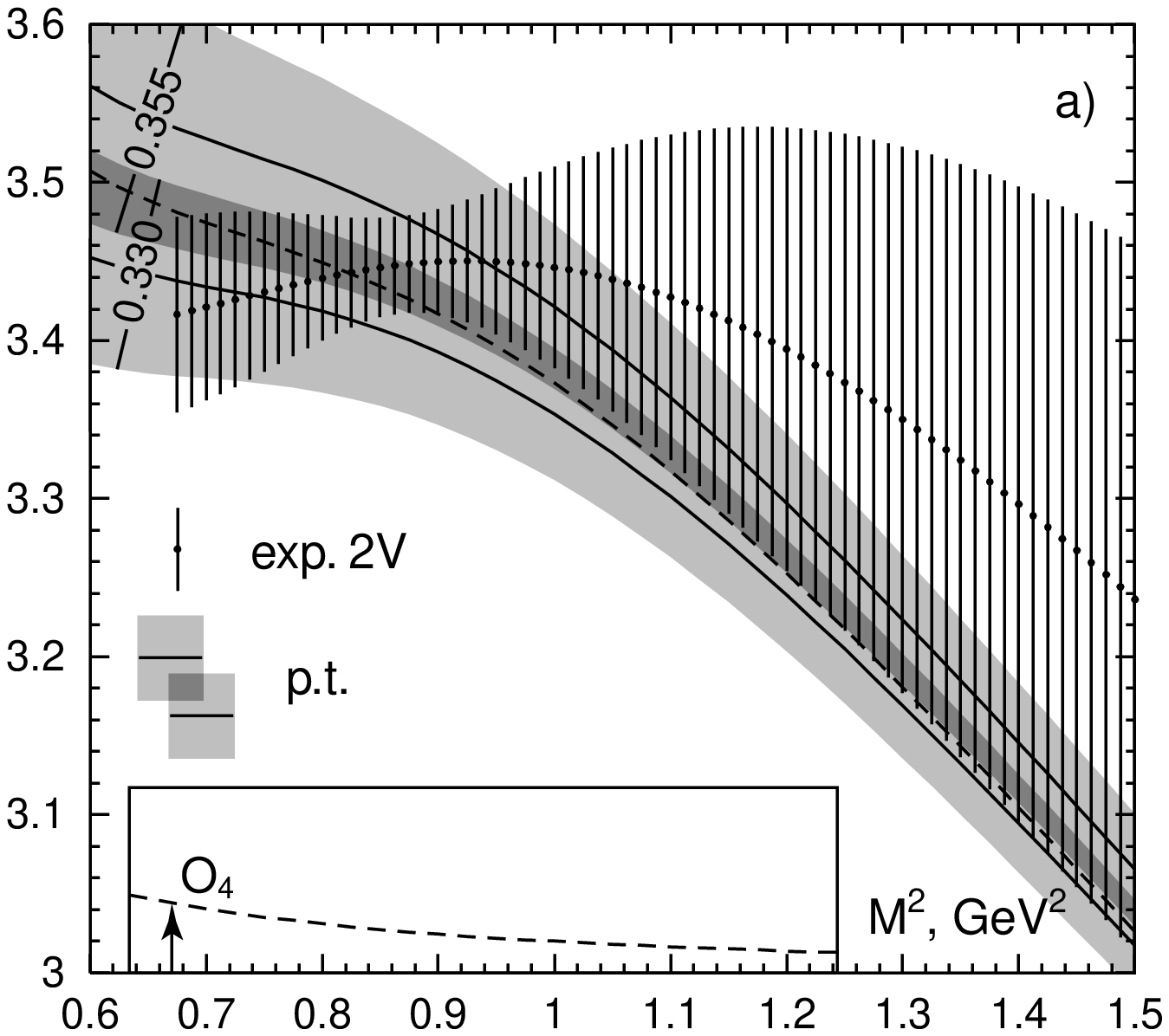, width=81mm}
\epsfig{file=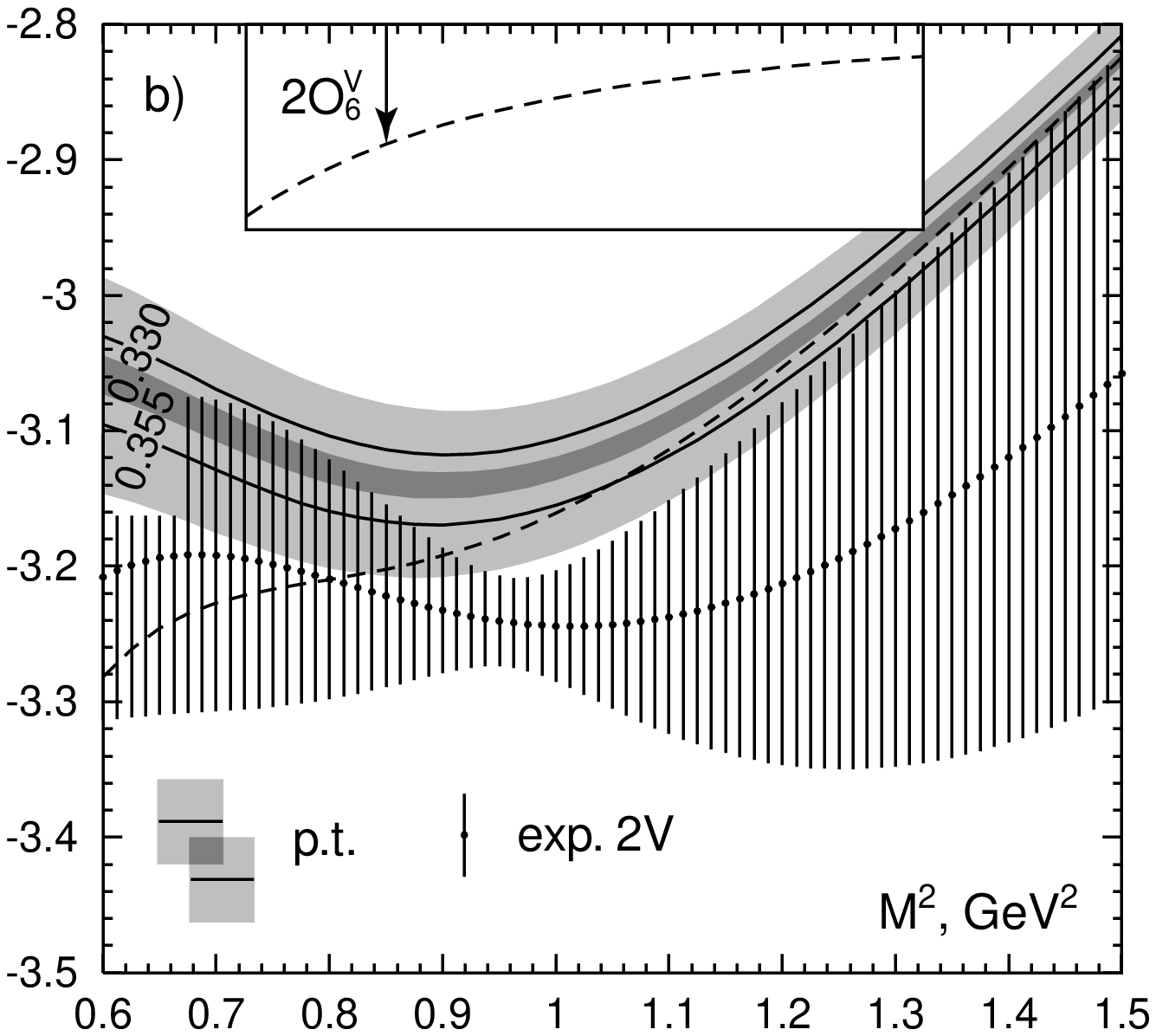, width=81mm}
\caption{The sum rules (\ref{set2}) (a) and (\ref{set3}) (b) for vector currents. }
\label{set23fig}
\end{figure}

At first we consider usual Borel transformation for vector current correlator, since it was
originally applied in \cite{EVK} for the sum rule analysis.
 It is defined as (\ref{bor0}) with the experimental
spectral function $\omega_{\rm exp}=2v_1$ instead of $v_1+a_1+a_0$
(the normalization is $v_1(s) \to 1/2$ at $s\to \infty$ in parton model). Respectively, in the
r.h.s. one should take the vector operators $2O^V=O^{V+A}+O^{V-A}$, all $O^{V-A}$ with
$D\le 8$ can be found in \cite{IZ}.   Numerical results are shown in Fig \ref{borvfig}.
The perturbative theoretical curves are the same as in Fig \ref{bor0fig} with $V+A$
correlator. The dashed lines display the contributions of the gluonic
condensate given by (\ref{condnum}), $2O_6^V=-5.5\times 10^{-3}\,{\rm GeV}^6$
and $2O_8^V=O_8^{V-A}=7\times 10^{-3} \,{\rm GeV}^8$
added to the $0.330$-perturbative curve. The contribution of each condensate is shown in the
box below. Notice, that for such condensate values the total OPE contribution
is small, since positive $O_4$ and $O_8$ compensate negative $O_6$.
The agreement is observed for $M^2>0.8\,{\rm GeV}^2$.

Now we apply the method of Borel transformation along the rays to the vector
polarization operator to separate the contribution of different operators from each other.
The $D=8$ operator is important here, so we shall separate $O_{4,6}$ from $O_8$.

Borel transformation at low $M^2$ exponentially suppresses the contribution
of large $s$ domain, where the experimental error is high. 
Besides of this we may use the oscillating behavior
of the complex exponent to further suppress the high-error points 
near $s=m_\tau^2$. This would allow us to go to higher $M^2$.
Here the real part of $B(M^2)$ has obvious advantage since
the function $\cos{(\phi + {s\over M^2}\cos{\phi})}$ has zero at $s=m_\tau^2$ and $\phi\sim \pi/4$
already at $M^2\approx 1\,{\rm GeV}^2$, while the largest (in $M^2$) zero 
of $\sin{(\phi + {s\over M^2}\cos{\phi})}$
in imaginary part is twice lower. So let us take three different angles, say,
$\phi=0, \pi/6, \pi/4$. Solving the system of linear equations, we get:
\bea
 {\rm Re} \, \left[ \,{B(0)\over 2-\sqrt{2}}
\,-\,\sqrt{2}\,B(\pi/6)\, + \,{B(\pi/4) \over \sqrt{2}-1} \,\right]
& = & {\rm p.t.} \,+ \,2\pi^2\,{O_4\over M^4}  \label{set2} \\
 {\rm Re} \, {-\,B(0)\,+\,2\,B(\pi/6)\, -\, 2\, B(\pi/4) \over \sqrt{2}-1} & = & {\rm p.t.} \,+
 \,2\pi^2\,{O_6\over 2 M^6}   \label{set3}
\eea
For brevity we write here $B(\phi)$ instead of $B_{\rm exp}(M^2e^{i\phi})$, "p.t." stands for
perturbative contribution.
The results for the equations (\ref{set2}, \ref{set3}) are
shown in Figs \ref{set23fig}a,b respectively.

Fig \ref{set23fig}a demonstrates, that the vector sum rule is satisfied at 
$\alpha_s(m_\tau^2)=0.330$ and gluonic condensate (\ref{condnum}) 
(although higher values of gluonic condensate, e.g.~SVZ value still
do not contradict the data). Fig \ref{set23fig}b shows, that the $O_6^V$ contribution works
in right direction: its addition to 0.330-perturbative curve shrinks
the disagreement between the theory and experiment. However, some 
discrepancy (about 0.04, i.e.~$0.1\%$ in the worst case)  still persist.
It may be addressed either to the uncertainty in $\alpha_s(m_\tau^2)$ ---
a slightly higher value would be desirable, or to the underestimation of
$O_6^V$ (in absolute value: $O_6^V$ is negative) by $20-30\%$, or both.
Remind, that the numerical values of condensates depend on the way, how the
infrared region is treated ($O_6$ is chirality conserving). We are considering
here 3-loop condensates, defined in Sec.~4. The $O_6^V$ value was
taken equal to $7/18$ of $O_6^{V-A}$, obtained from $V-A$ data analysis
\cite{IZ}, where perturbative terms are absent, and some difference is not 
excluded.

\section{The check for renormalon-type terms}

In the asymptotic perturbative series a special part of terms --- renormalons (infrared and ultraviolet)
is often separated and the summation of them is performed (for a recent review see
\cite{BB}). In such sum the term appears proportional to $1/Q^2$ at large $Q^2$ and looking as a
contribution of $D=2$ operator. (In OPE the $D=2$ operator is proportional to $m_q^2$ and is
very small.) Renormalons conserve chirality and may contribute to $V+A$ but not to $V-A$.
Unfortunately, the coefficient in front of the $1/Q^2$ term of the renormalon origin cannot be
calculated reliably. (In \cite{Susl} it was claimed, that the renormalons are totally absent in the
perturbative series asymptotics and therefore this coefficient is zero.) In recent paper \cite{CNZ}
the hypothesis was suggested, that infrared renormalons result in substitution
\be
{\alpha_s\over \pi} \, \to \,{\alpha_s\over \pi} \left(\,1\,-\,1.05\,{\lambda^2\over Q^2}  \right)
\label{tach}
\ee
in the first $\alpha_s$ correction to polarization operator or Adler function (the $Q^2$-dependence of
$\alpha_s$ was not accounted in \cite{CNZ}).  In (\ref{tach}) $\lambda^2$ is tachionic gluon mass,
$\lambda^2<0$ and for its value the estimation was found:
\be
-\,\lambda^2\,=\, (0.2 - 0.5) \, {\rm GeV}^2
\label{lamtach}
\ee
The authors of \cite{CNZ} could not discriminate even the highest value $\lambda^2=-0.5\,{\rm GeV}^2$.

\begin{figure}[tb]
\hspace{30mm}
\epsfig{file=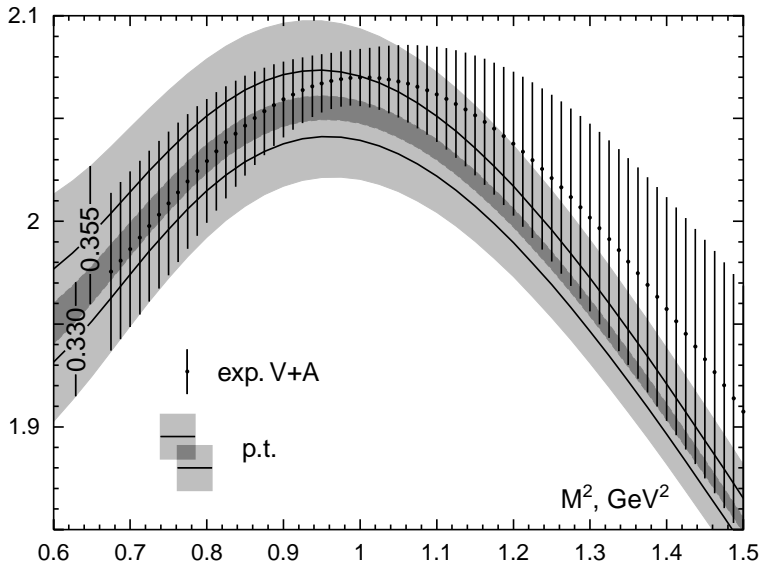, width=100mm}
\caption{Sum rule (\ref{set10}) with $O_2$ but without $O_{4,6}$}
\label{set10fig}
\end{figure}

Let us try to find the restriction on $O_2$ operator  from the sum rule for $V+A$ correlator in the
complex $q^2$-plane from ALEPH data. (We call it for brevity $O_2$, although it is
not the $D=2$ operator which stands in OPE.) As we did in previous section, for this
purpose we take the real part of the Borel transform (\ref{bor0}) $B(M^2e^{i\phi})$ at the
angles $\phi=0,\pi/6, \pi/4$ and separate the operator $O_2$ from $O_{4,6}$:
\be
 {\rm Re} \, {B(0)\,-\,2\,B(\pi/6)\, + \,\sqrt{2}\, B(\pi/4) \over 2-\sqrt{3}} \, = \, {\rm p.t.} \,+
 \,2\pi^2\,{O_2\over M^2}  \label{set10}
\ee
The experimental
and perturbative parts of this combination are plotted in Fig \ref{set10fig}.

The sum rule (\ref{set10}) shown in Fig \ref{set10fig} gives the following value of the dimension
2 operator:
\be
O_2\,=\,(1.0\pm 1.5)\times 10^{-3} \, {\rm GeV}^2  \; , \qquad  \alpha_s (m_\tau^2)=0.33
\label{o2lim}
\ee
We got this estimation at $M^2=1 \, {\rm GeV}^2$, where experimental error is minimal.  In the
model of \cite{CNZ}
\be
 O_2\,=\,-1.05\,{\alpha_s\over \pi}\,\lambda^2
\ee
At $\alpha_s(1\,{\rm GeV}^2)/\pi=0.18$, corresponding to $\alpha_s(m_\tau^2)=0.33$,
 there follows the restriction from (\ref{o2lim}):
\be
-\,\lambda^2\,=\,(5\pm 8)\times 10^{-2} \, {\rm GeV}^2
\ee
which is few times smaller than even the lower limit in (\ref{lamtach}). Notice, that similar restrictions
on the value of $D=2$ operator have been obtained in \cite{Dom} from other sum rules.

\section{Instanton corrections}

Some nonperturbative features of QCD may be described in so called instanton gas model
(see \cite{SShur} for extensive review and the collection of related papers in \cite{S2}).
Namely, one computes the correlators in the
$SU(2)$-instanton field embedded in the $SU(3)$ color group. In particular, the 2-point correlator
of the vector currents has been computed long ago \cite{AG}.
Apart from usual tree-level correlator $\sim \ln {Q^2}$ it has a correction which depends on
the instanton position and radius $\rho$. In the instanton gas model these parameters
are integrated out. The radius is averaged over some concentration $n(\rho)$, for which one
or another model is used. Concerning the
2-point correlator of charged axial currents, the only difference from the vector case is that
the term with 0-modes must be taken with opposite sign. In coordinate representation the
answer can be expressed in terms of elementary functions, see \cite{AG}. An attempt to
compare the instanton correlators with ALEPH data in coordinate space has been undertaken
in \cite{SShur2}.

We shall work in momentum space. Here the instanton correction to the spin-$J$
parts $\Pi^{(J)}$ of the correlator (\ref{pol0}) can be written in the following  form:
\bea
\Pi^{(1)}_{V,\,{\rm inst}}(q^2) & =& \int_0^\infty\!d\rho \, n(\rho)\, \left[\,
-\,{4\over 3 q^4}\,+\,\sqrt{\pi}\rho^4 G^{30}_{13}
\left( -\rho^2 q^2 \left| { 1/2 \atop 0,0,-2 }\right. \right) \right]
  \nonumber \\
\Pi^{(0)}_{A,\,{\rm inst}}(q^2) & =& \int_0^\infty\!d\rho \, n(\rho)\, \left[ \,-\,{4\over q^4}\, -
\,{4\rho^2\over q^2}\,K_1^2\!\left(\rho\sqrt{-q^2}\right) \right] \nonumber \\
\Pi^{(1)}_{A,\, {\rm inst}}(q^2) & = &
\Pi^{(1)}_{V,\, {\rm inst}}(q^2)-\Pi^{(0)}_{A,\, {\rm inst}}(q^2) \; , \qquad
\Pi^{(0)}_{V,\, {\rm inst}}(q^2)\,=\,0
\label{pif}
\eea
Here $K_1$ is modified Bessel function, $G_{mn}^{\,p\,q}(z|\ldots)$ is Meijer function.
Definitions, properties and approximations of Meijer functions can be found, for instance,
in \cite{Luke}. In particular the function in (\ref{pif}) can be written as the following series:
\bea
\sqrt{\pi}G^{30}_{13}\left( z\left| { 1/2 \atop 0,0,-2 }\right. \right) &=& {4\over 3z^2}\,-\,{2\over z}\,
+\,{1\over 2\sqrt{\pi}}\,\sum_{k=0}^\infty \,z^k {\Gamma(k+1/2)\over \Gamma^2(k+1)\,\Gamma(k+3)}\times
 \nonumber \\
 & &
\times\Bigl\{\, \left[\, \ln{z}\,+\,\psi(k+1/2)\,-\,2\,\psi(k+1)\,-\,\psi(k+3)\, \right]^2 \nonumber \\
 & &+\,\psi'(k+1/2)\,-\,2\,\psi'(k+1)\,-\,\psi'(k+3) \,\Bigr\} \label{meijer}
\eea
where $\psi(z)=\Gamma'(z)/\Gamma(z)$. For large $|z|$ one can obtain its approximation by
the saddle-point method:
\be
G^{30}_{13}\left( z\left| { 1/2 \atop 0,0,-2 }\right. \right) \approx
\sqrt{\pi} z^{-3/2} e^{-2\sqrt{z}} \; , \qquad |z|\gg 1
\label{meijas}
\ee

The formulas (\ref{pif}) should be treated in the following way. One adds $\Pi_{\rm inst}$ to usual
polarization operator (\ref{polop1}) with perturbative and OPE terms. But the terms $\sim 1/q^4$
\underline{must be absorbed} by the operator $O_4$ in (\ref{polop1}), since the gluonic condensate
$\bigl<G^2\bigr>$ is averaged over all field configurations, including
the instanton one. Notice negative sign before $1/q^4$ in (\ref{pif}). It happens because
the negative contribution of the quark condensate $\bigl<m\bar{q}q\bigr>$ in the instanton field
exceeds positive contribution of the gluonic condensate $\bigl<G^2\bigr>$.
In real world $\bigl<m\bar{q}q\bigr>$ is negligible at $q^2 \sim 1 \, {\rm GeV}^2$.

The correlators (\ref{pif}) possess appropriate analytical properties, they have a cut along positive
real axes:
\bea
{\rm Im}\,\Pi^{(1)}_{V,\, {\rm inst}}(q^2+i0) & =& \int_0^\infty\!d\rho \, n(\rho)\,\pi^{3/2} \rho^4 G^{20}_{13}
\left(\rho^2 q^2 \left| { 1/2 \atop 0,0,-2 }\right. \right)
\label{repif} \\
 {\rm Im}\,\Pi^{(0)}_{A,\, {\rm inst}}(q^2+i0)& = &-\,\int_0^\infty\!d\rho \, n(\rho)\,
{2\pi^2\rho^2\over q^2} J_1\!\left(\rho\sqrt{q^2}\right)N_1\!\left(\rho\sqrt{q^2}\right)
\label{impif}
\eea

We shall consider below the instanton concentration advocated by Shuryak (see \cite{SShur}
and references therein). It is a model with fixed instanton radius
(RILM model in \cite{SShur}):
\be
n(\rho)\,=\,n_0 \, \delta(\rho - \rho_0)
\ee
 From \cite{SShur} we take the numbers:
\be
\rho_0\,=\,1/3\,{\rm fm} \, = \,1.7 \, {\rm GeV}^{-1} \; , \qquad
n_0\,=\,1\,{\rm fm}^{-4} \, = \,1.5 \times 10^{-3} \, {\rm GeV}^4
\label{n0rho0}
\ee

Now we consider the instanton contribution to the $\tau$-decay branching ratio (\ref{rtau2}).
Since the instanton correlator (\ref{pif}) has $1/q^2$ singular term in the expansion
near 0 (see (\ref{meijer})), the integrals must be taken over the circle, like in (\ref{rtau3}).
In the instanton model the function $a_0(s)$ differs from experimental $\delta$-function, which gives
small correction (\ref{delr0}). So we shall ignore the last term in (\ref{rtau2}) and consider the
integral with $\Pi_{V+A}^{(1)}+\Pi_A^{(0)}$
in (\ref{rtau3}). Here we need the following formulas for the circle integrals, which can be rigorously
obtained from the series representation of the Meijer function (\ref{meijer}):
\bea
{i\over 2\pi} \oint_{|s|=s_0} {ds\over s_0} \left( {s\over s_0} \right)^k
G^{30}_{13}\left( -\rho^2 s\left| { 1/2 \atop 0,0,-2 }\right. \right) & = &
G^{21}_{24}\left( \rho^2 s_0\left| { -k,\,1/2 \atop 0,0,-2,-k-1 }\right. \right) \; , \qquad k\ge 2
\nonumber   \\
{i\over 2\pi} \oint_{|s|=s_0} {ds\over s_0}\, {s\over s_0} \,
G^{30}_{13}\left( -\rho^2 s\left| { 1/2 \atop 0,0,-2 }\right. \right) & =  &
-{4\over 3\sqrt{\pi} \rho^4s_0^2} +
G^{21}_{24}\left( \rho^2 s_0\left| { -1,1/2 \atop 0,0,-2,-2 }\right. \right)
\nonumber \\
{i\over 2\pi} \oint_{|s|=s_0} {ds\over s_0} \,
G^{30}_{13}\left( -\rho^2 s\left| { 1/2 \atop 0,0,-2 }\right. \right)  & =  &
-G^{20}_{13}\left( \rho^2 s_0\left| { 1/2 \atop 0,-1,-2 }\right. \right) \label{meijint}
\eea
The first term in the r.h.s.~of the second equation looks like the
contribution of $D=4$ operator, but in fact it is not. Indeed,
all expressions in the r.h.s.~of (\ref{meijint}) have the same LO term 
of the asymptotic expansion for large $s_0$, equal to 
$-\sin{(2\rho\sqrt{s_0})}/(\sqrt{\pi}\rho^4s_0^2)$. However
for $k\ge 3$ the accuracy of this approximation is bad and exact values 
of Meijer functions should be used for numerical evaluations.

\begin{figure}[tb]
\epsfig{file=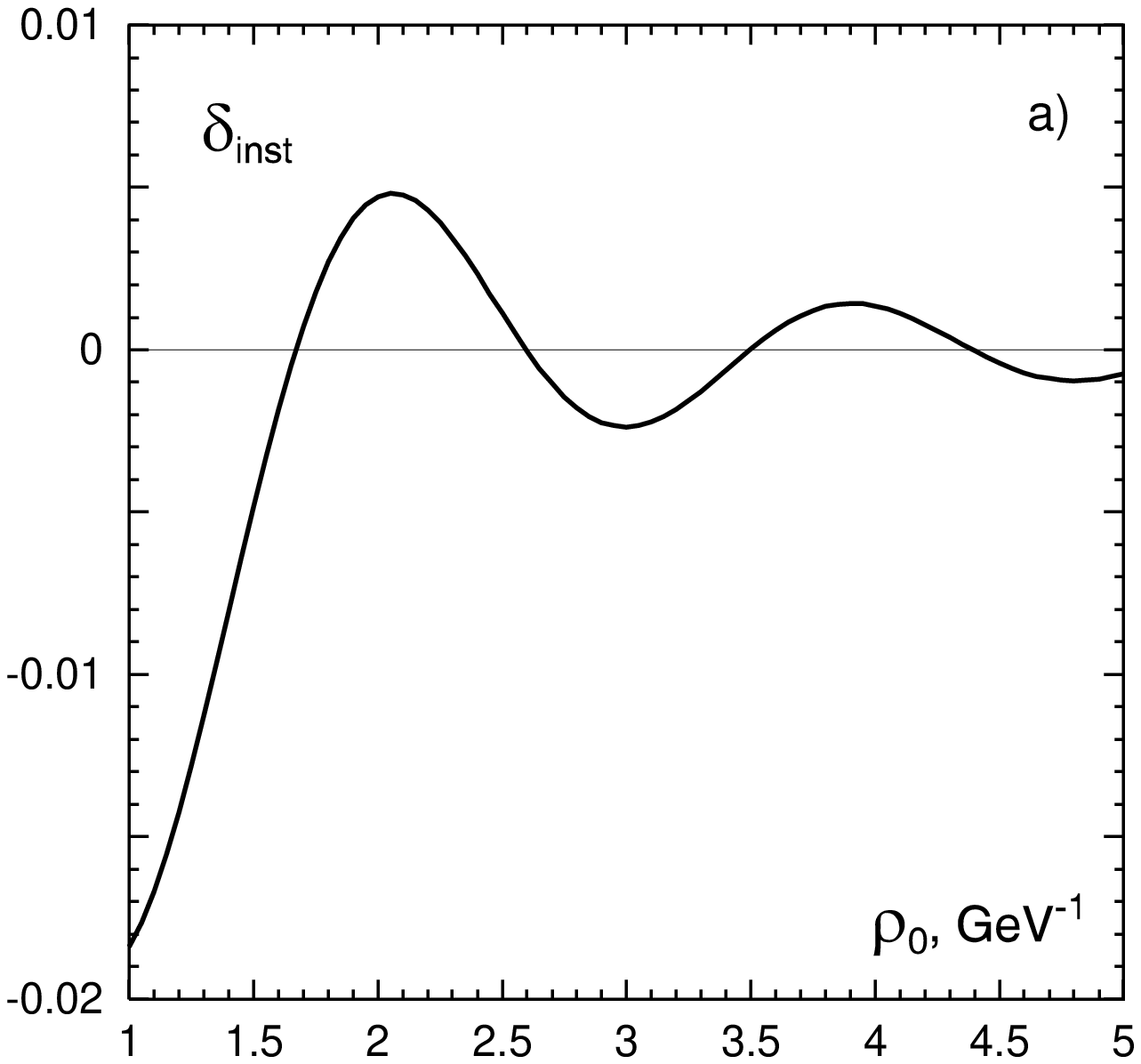, width=81mm}
\epsfig{file=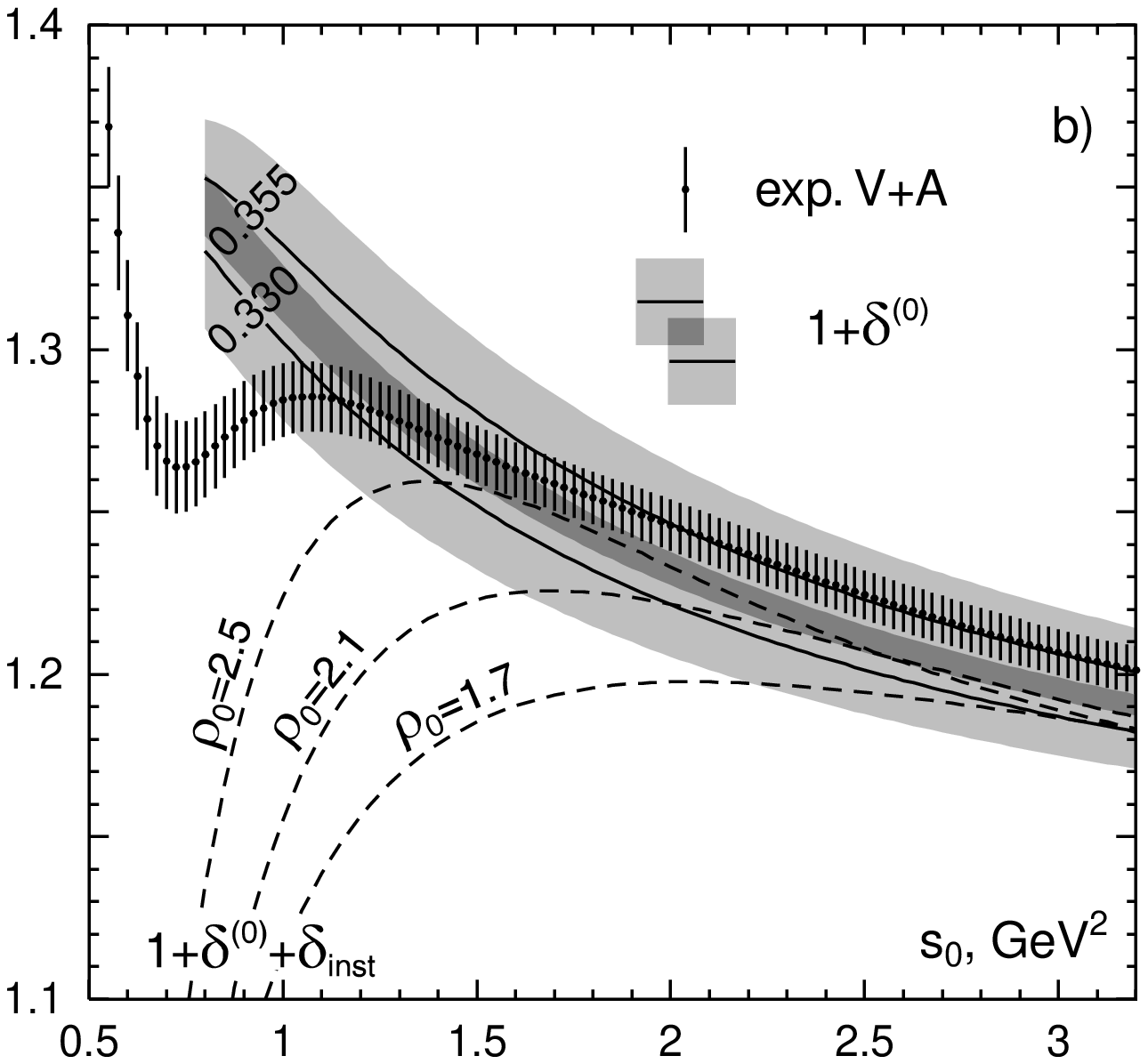, width=81mm}
\caption{The instanton correction to the $\tau$ decay ratio versus $\rho_0$ (a) and
 "versus $\tau$ mass" (b) for $n_0=1.5 \times 10^{-3}\,{\rm GeV}^4$ }
\label{inst_tau}
\end{figure}

With help of (\ref{meijint}) the instanton correction to the $\tau$-decay branching ratio
can be brought to the following form:
\be
\delta_{\rm inst}\,  =\,  -\,48\,\pi^{5/2}\int_0^\infty\!d\rho \, n(\rho)\,\rho^4\,
G^{20}_{13}\left(\rho^2 m_\tau^2 \left| { 1/2 \atop 0,-1,-4 }\right. \right) \approx \,
{48\pi^2 n_0 \over \rho_0^2 m_\tau^6} \, \sin{(2\rho_0 m_\tau)}
\label{dtinst}
\ee
Since the parameters (\ref{n0rho0}) are determined quite approximately, we may
explore the dependence of $\delta_{\rm inst}$ on them. The $\delta_{\rm inst}$ versus
$\rho_0$ for fixed $n_0$ (\ref{n0rho0}) is shown in Fig \ref{inst_tau}a.

As seen from Fig \ref{inst_tau}a the instanton correction to hadronic $\tau$-decay is extremely small
except for unreliably low value of the instanton radius $\rho_0<1.5\,{\rm GeV}^{-1}$. At the favorable
value \cite{SShur} $\rho_0=1.7\,{\rm GeV}^{-1}$ the instanton correction to $R_\tau$ is almost
exactly zero. (Of course, smaller values of $n_0$ than (\ref{n0rho0}) are also allowed.)
This fact confirms our calculations of $\alpha_s(m_\tau^2)$ (Sec.~2), where the
instanton corrections were not taken into account.

The result (\ref{dtinst}) can be used in another way. Namely, the $\tau$ mass can be considered
as free parameter $s_0$.
The dependence of the fractional corrections $\delta^{(0)}$ and
$\delta^{(0)}_{0.330}+\delta_{\rm inst}$
on $s_0$ is shown in Fig \ref{inst_tau}b\footnote{The Fig \ref{inst_tau}b can be
compared with Figure 15 in ALEPH paper \cite{ALEPH2}. The discrepancy between
theoretical curves at $s_0<1\,{\rm GeV}^2$ is explained by different approximations: we used
3-loop perturbation theory, while the authors of \cite{ALEPH2} used 4-loop one with $K_4=50\pm 50$.}.
The result strongly depends on the instanton radius and rather essentially
on the density $n_0$. For $\rho_0=1.7\,{\rm GeV}^{-1}$ and 
$n_0=1\,{\rm fm}^{-4}$ (\ref{n0rho0}) the instanton curve is outside
the errors already at $s_0\sim 2\,{\rm GeV}^2$, where the perturbation theory 
is expected to work. Therefore Fig \ref{inst_tau}b 
shows, that in RILM model the instanton
radius must be larger (say, $\rho_0=2.5\,{\rm GeV}^{-1}$) or the instanton
density much ($2-3$ times) lower. The contribution of $D=6$ operator
$\delta_{V+A}^{(6)}$ is not shown on Fig \ref{inst_tau}b. It is equal
$\delta_{V+A}^{(6)}=-(5\pm 2)\times 10^{-3} (m_\tau^6/s_0)^3$ and 
quite large at $s_0<1.5\,{\rm GeV}^2$. 

Consequently in this approach
the perturbation theory + OPE + RILM (at not very large $\rho_0$) cannot
satisfactory describe the data at $s_0<1.5\,{\rm GeV}^2$. Since the
instanton contribution is large here, we disbelieve all the
results, obtained by the method of variable $\tau$-mass in this domain.
(Perhaps, the shadowed region in Fig \ref{pt_ope} is of importance in this
method at low $s_0$.)

The $\tau$ decay ratio is not sensitive to the gluonic condensate.
Let us consider now the sum rules which depend on  it. The Borel transformation
of the instanton part is:
\bea
{\cal B}_{M^2} \Pi_{\rm inst} & = &
2\pi i \oint e^{-s/M^2} \,\Pi_{V,\, {\rm inst}}^{(1)}(s)\,{ds\over M^2} \nonumber \\
 &=&4\pi^2 \int_0^\infty\!d\rho \, n(\rho)\,\left[\,-\,{4\over 3 M^4}\,+\,\sqrt{\pi}\rho^4 G^{20}_{12}
\left(\rho^2 M^2 \left| { 1/2 \atop 0,-2 }\right. \right) \right]
\label{binst}
\eea
The integration contour goes around the cut from 
$s=+\infty +i0$ to $s=+\infty-i0$. 
The term $\sim 1/M^4$ here comes from the term $\sim 1/q^4$ in
(\ref{pif}); it must be included in the $\bigl<O_4\bigr>$ contribution
in (\ref{bor0}). The Meijer function in (\ref{binst}) has the asymptotics
$$
G^{20}_{12}\left( z \left| { 1/2 \atop 0,-2 }\right. \right)\,  \approx \, z^{-5/2} e^{-z} \; , \qquad |z|\gg 1
$$
and strongly suppressed at $M^2>0.8\,{\rm GeV}^2$. We calculated the instanton contribution
to all Borel-like sum rules used here; it is indeed negligible compared to the errors. Consequently the
results of previous sections remain unchanged.

\begin{figure}[tb]
\hspace{35mm}
\epsfig{file=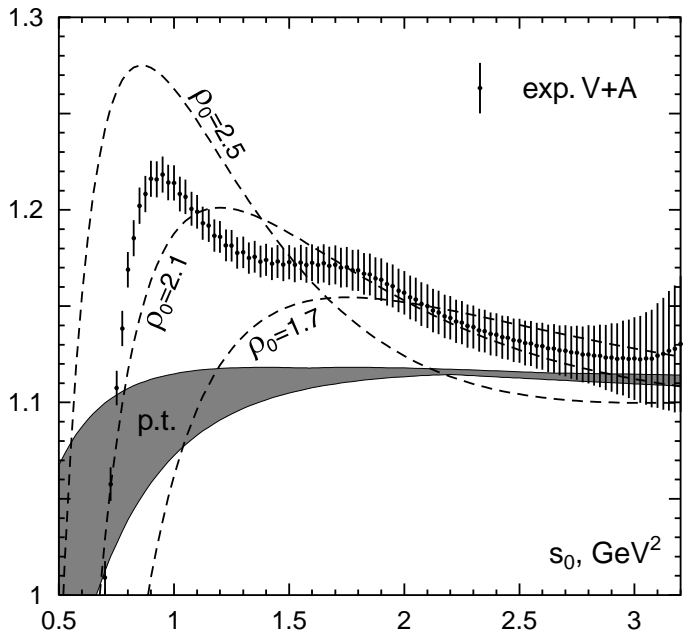, width=90mm}
\caption{Sum rule (\ref{mom3}). Experimental, pure perturbative and "perturbative + instanton"
parts are shown. The $O_4$ contribution is not taken into account.}
\label{mom3fig}
\end{figure}

However the spectral moments sum rules, often used in $\tau$-decay data analysis \cite{ALEPH2},
can be quite sensitive to the instanton corrections. Let us consider the following sum rule,
constructed in this way:
$$
4\int_0^{s_0} {ds\over s_0}\,{s\over s_0} \left(1-{s^2\over s_0^2}\right) \omega_{\rm exp} (s) \,=\,
{\rm p.t.} \,-\, 8\pi^2 {\bigl<O_4\bigr> \over s_0^2}
\,+\, 16\pi^2 \int_0^\infty\!d\rho \, n(\rho)\,\rho^4\left[ \,-{4\over 3\rho^4s_0^2}\right. \hspace{15mm}
$$
\be \hspace{15mm}
 \left. +
\sqrt{\pi} G^{21}_{24}\left(\rho^2 s_0 \left| { -1,1/2 \atop 0,0,-2,-2 }\right. \right) -
\sqrt{\pi} G^{21}_{24}\left(\rho^2 s_0 \left| { -3,1/2 \atop 0,0,-2,-4 }\right. \right)
 \right]
\label{mom3}
\ee
The integral (\ref{mom3}) is normalized to 1 in parton model. 
It does not depend on the $D=6$ operator, the 
factor $1-s^2/s_0^2$ is introduced to suppress
large experimental errors for large $s_0$. Remind our convention:
the contribution of the term $\sim 1/q^4$ in $\Pi_{\rm inst}$
(\ref{pif}) is included into 
the operator $\bigl<O_4\bigr>$ in (\ref{mom3}).
The contribution of different parts of eq.~(\ref{mom3})
are plotted versus $s_0$ in Fig \ref{mom3fig}. Since the weight function in the integral vanishes
at $s=0$, the contribution of unphysical cut is suppressed. So the theoretical errors are
diminished here as well as the sensitivity on various perturbative parameters. The
theoretical curve is shown as single shaded area, which includes both the
uncertainty of $\alpha_s(m_\tau^2)$ and the error $\pm K_3a^3$ for each $\alpha_s(m_\tau^2)$.

The operator $\bigl<O_4\bigr>$ enters with negative sign in (\ref{mom3}), so the theoretical curve
must go above experimental one. This is certainly not the case if the instanton corrections
are not taken into account. For $\rho_0=2.1\,{\rm GeV}^2$ the theoretical and experimental
results are in good agreement for $\bigl<O_4\bigr>=0$. 
By increasing the instanton density $n_0$, 
positive values of $\bigl<O_4\bigr>$ become possible. In this aspect sum 
rules (\ref{dtinst}) with varying $m_\tau$ and (\ref{mom3}) are not 
in agreement:  (\ref{dtinst}) favors small $n_0$ 
while (\ref{mom3}) prefers large $n_0$. 

These results are, however, not convincing. The main conclusion, coming
from consideration of spectral moments sum rules, is that they are not
suitable for QCD analysis  untill we have a complete theory. 
(This statement surely refers also to the method, where $\tau$-mass
is considered as free parameter.) The same situation took place for
$V-A$ correlators: the spectral moments sum rules worked only
at the circle radius $s_0>2\,{\rm GeV}^2$ \cite{IZ}.

\section{Conclusion}

The goal of this paper was to confront the recent precise experimental data on hadronic $\tau$-decay
with QCD calculations at low $Q^2$ and to check the basic aspects of QCD: perturbative series,
OPE as well as various nonperturbative QCD approaches. The data present the imaginary part of
polarization operators ${\rm Im}\,\Pi_{V,A}(s)$, $s=q^2$ at $0<s<m_\tau^2$. If some procedure is
applied to suppress or nullify the influence of high energy domain  (Borel transformation,
integration over closed circle in complex $s$-plane), then with help of dispersion relation
the values of $\Pi_{V,A}(s)$ is the whole complex $s$-plane at low $|s|$ can be found from experiment.
(By low $|s|$ we mean $|s|<2-3\,{\rm GeV}^2$.) These experimental values of $\Pi_{V,A}(s)$
can be compared with theoretical calculations in the domain of complex $s$-plane, where QCD
describes the data well enough, in order to find the values of QCD parameters: $\alpha_s$ and
condensates.

In \cite{IZ} this program was
 realized for $\Pi_{V-A}$ polarization operator and the values of dimension
6 and 8 condensates were found. In this paper $\Pi_{V+A}$ and $\Pi_V$ polarization operators
were studied, where perturbative contribution is dominant (unlike $\Pi_{V-A}$ which is given entirely
by condensates). It must be stressed, that the present situation has changed drastically in
comparison with earlier study of similar problem \cite{EVK}. In \cite{EVK} the perturbative contribution
was much less essential and the authors could restrict theirselves to LO term only. In this paper
the perturbative calculations were performed in 3 and 4 loop
approximation. The unphysical cut in the complex $s$-plane in perturbative part of the polarization
operator was taken into account and the calculations (at least partly) were performed in such a way,
which allows one to minimize its influence (e.g. the Borel transformation along the rays, going
from the origin at some angle). The terms of OPE were accounted up to dimension $D=8$. It was
shown that $D=8$ contribution is very small in case of $V+A$ correlator. The coincidence of theoretical
and experimental values with accuracy better than $2\%$ was required. Let us remind that usually
the accuracy of standard QCD sum rule calculations is of order $10-15\%$.

The following results have been obtained:
\begin{enumerate}
\item
The value of QCD coupling constant $\alpha_s(m_\tau^2)=0.355\pm 0.025$ was found from hadronic
branching ratio $R_\tau$. It was shown, the sum rules at low $|s|$ favor the value close to the
lower error edge $\alpha_s(m_\tau^2)=0.330$ corresponding to $\alpha_s(m_Z^2)=0.118$.
\item
It was demonstrated that QCD with inclusion of OPE terms is in agreement with the data at
the values of complex Borel parameter $|M^2|>0.8 -1.0 \,{\rm GeV}^2$ in the left complex half-plane.
\item
The restriction on the value of the gluonic condensate was found
$\bigl<{\alpha_s\over\pi}G^2\bigr> = (0.006\pm 0.012)\, {\rm GeV}^4$ in comparison with standard
SVZ value $0.012\,{\rm GeV}^4$.
\item
The value of $D=6$ condensate found in \cite{IZ} is in agreement with $V+A$ and $V$ sum rules,
but cannot be specified.
\item
The analytical perturbative QCD \cite{ShSol1, ShSol3, GI} was compared with the data and it was
demonstrated that this approach is in strong contradiction with experimental value of $R_\tau$.
\item
The restrictions on $1/Q^2$ term in polarization operator of renormalon origin were found,
much stronger, than in the previous investigation \cite{CNZ}.
\item
The instanton contributions to polarization operator were analyzed and compared with the data in
the framework of the random instanton liquid model (RILM) \cite{SShur}. It was shown that the
instanton contribution to $R_\tau$ is very small, the same is true for Borel sum rules.
However their contributions can be significant to the spectral moments sum rules, often
used in $\tau$-decay data analysis.
\item
It was found that the method of spectral moments (integration over 
the circle with a polynomial) is less effective in the study of
the polarization operators at low $Q^2$, than Borel sum rule because of
larger contribution not given by OPE nonperturbative corrections
(see Sec.~7 and \cite{IZ}).
\end{enumerate}

We believe, that the results of this paper will serve for improving the QCD sum rules method.

\section*{Acknowledgement}

We are very thankful to M. Shifman, who paid our attention to his lectures \cite{S1}, for his valuable
correspondence about the problem under consideration and for his interest to our work.
B.I. thanks D.V. Shirkov for providing with exhaustive information about the
publications on analytical QCD, as well as for moral support.
B.I. is also thankful to J. Speth and N. Nikolaev for their hospitality at
Juelich FZ, where this work was finished.  The authors are indebted to M. Davier for
his kind presenting the ALEPH experimental data. We are indebted to G. Cvetic and T. Lee for
the remark which allowed us to find a misprint in previous version of the paper.

The research described in this publication was made possible in part by Award No RP2-2247
of U.S. Civilian Research and Development Foundation for Independent States of Former
Soviet Union (CRDF), by the Russian Found of Basic Research grant 00-02-17808 and
INTAS Call 2000, project 587.

\end{document}